\documentclass[pra,twocolumn,
showpacs,superscriptaddress,
,preprintnumbers,amsmath,amssymb,nofootinbib]{revtex4}
\usepackage{graphicx}
\usepackage{dcolumn}
\usepackage{bm}
\usepackage{slashed}
\usepackage{ulem} 
\usepackage{color}

\newcommand{\vek}[1]{\bm{\mathrm{#1}}}
\newcommand{\eq}{\mathit{eq}}
\newcommand{\coll}{\mathit{coll}}
\newcommand{\boxx}{\mathit{trap}}

\newcommand{\diff}{\mathit{diff}}
\newcommand{\sw}{\mathit{sw}}
\newcommand{\rms}{\mathit{rms}}
\newcommand{\ho}{\mathit{ho}}

\newcommand{\inter}{\mathit{int}}
\newcommand{\Eq}[1]{Eq.\@ (\ref{#1})}
\newcommand{\Eqs}[1]{Eqs.\@ (\ref{#1})}

\newcommand{\Fig}[1]{Fig.\@ \ref{#1}}

\newcommand{\Sec}[1]{Sec.\@ \ref{#1}}

\newcommand{\erfB}{\textrm{erf}(\sqrt{\beta V_0})}
\newcommand{\erfcB}{\textrm{erfc}(\sqrt{\beta V_0})}
\newcommand{\erf}{\textrm{erf}}
\newcommand{\erfc}{\textrm{erfc}}

\begin{document}
\title{Boltzmann equation with double-well potentials}

\author{Silvia Chiacchiera}
\affiliation{CFisUC, Department of Physics, 
University of Coimbra, P-3004-516 Coimbra, Portugal}

\author{Tommaso Macr\`i}
\affiliation{Departamento de F\'isica Te\'orica e Experimental,
Universidade Federal do Rio Grande do Norte, 59072-970 Natal-RN,Brazil}
\affiliation{International Institute of Physics, 59078-400 Natal-RN, Brazil}

\author{Andrea Trombettoni}
\affiliation{CNR-IOM DEMOCRITOS Simulation Center, Via Bonomea 265, 
I-34136 Trieste, Italy}
\affiliation{SISSA and INFN, Sezione di Trieste, Via Bonomea 265, I-34136 
Trieste, Italy}

\begin{abstract}
We study the dynamics of an interacting classical gas trapped 
in a double-well potential at finite temperature.
Two model potentials are considered: a cubic box with a square 
barrier in the middle, and a harmonic trap with a gaussian barrier 
along one direction.
The study is performed using the Boltzmann equation, solved 
numerically via the test-particle method. 
We introduce and discuss a simple analytical model 
that allows to provide estimates of the relaxation time, 
which are compared with numerical results. 
Finally, we use our findings to make numerical and analytical predictions 
for the case of a fermionic mixture in the normal-fluid phase in a realistic 
double-well potential relevant for experiments with cold atoms.
\end{abstract}
\maketitle

\section{Introduction}

Double-well energy potentials, with two, degenerate or not, 
minima separated by a maximum, are ubiquitous in several branches of science. 
These potentials are used to model a variety of processes involving an energy 
barrier \cite{Hanggi90}, 
from the computation of rate coefficients in a chemical reaction 
\cite{Calef83} and the modeling of solid-state junctions \cite{Barone82}
to the calculations involving non perturbative instanton configurations in 
quantum field theory \cite{Rajaraman87}. The importance \textit{per se} of 
the study of the dynamics 
in double-wells potentials stems also from the fact that it is often 
preliminary to (and useful for) the investigation of 
more complex dynamical effects in multi-well potentials.

The prototypical problem of the dynamics in a double-well potential is to 
determine in how much time the particles move from one well 
to the other one and if (and in how much time) 
they equilibrate reaching a vanishing $\Delta N$, where $\Delta N$ is 
the difference of population between the two wells. A source  
of inter-well motion is of course given by the quantum tunneling 
\cite{Barone82,Razavy03}. 
When quantum effects are dominant, a single particle tunnels 
from one well to the other, and for many particles the macroscopic 
quantum coherence exhibited at low temperature by many systems 
-- including $^4$He and $^3$He, ultracold atoms and superconductors 
\cite{Tilley,Annett} -- 
allows for a net current between the wells. At variance, at high temperatures 
thermal effects give rise to noise-assisted hopping events \cite{Hanggi90}: 
for ultracold atoms, at temperatures higher than the temperature at 
which the effects of quantum statistics become relevant, this 
corresponds to an incoherent flow of atoms and the ensuing thermalization 
of the two-well systems with $\Delta N \to 0$ \cite{macri13}. 

Experiments with cold atoms in double-well potentials give the 
concrete possibility 
to explore physical situations in which both phenomena -- quantum tunneling 
and thermal noise-assisted hoppings -- are present. 
The high degree of control of atomic gases at very low temperatures   
\cite{lewensteinbook} allows 
to design and perform experiments where particles, 
either bosons, fermions or mixtures of both, are subjected to
properly engineered and highly tunable trapping potentials. 
In this context, superfluid double- and multi-well dynamics 
have been extensively studied, both theoretically 
and experimentally, primarily for bosonic atoms and more recently also 
for fermionic gases. For bosons in the superfluid regime at $T=0$, 
which are well described by the Gross-Pitaevskii 
equation, this has lead 
to the identification of the atomic analog of the Josephson effect of 
superconductivity and of the macroscopic quantum 
self-trapping effect \cite{Sme97,Alb05,Ank05}, 
a direct consequence of the nonlinearity of the dynamical equations of motion. 
The study of superfluid fermions in double-well potentials is more recent and 
experiments for these systems are in progress.  
The peculiarity of such fermionic systems is, 
among others, the tunability of the strength of the 
inter-particle interactions via the so-called Feshbach resonances, 
which results in the well known
BCS-BEC crossover \cite{Zwerger}. 
In a recent experiment \cite{Valtolina15} Valtolina \textit{et al.} 
studied ultracold fermionic $^{6}$Li atoms in two different hyperfine states
loaded in double-well potentials, reporting on 
the observation of the Josephson effect between fermionic superfluids along the crossover.

Regarding thermal effects, 
another class of experiments focused during the years 
on the study of \textit{i)} polarized and two-component fermionic 
gases across and above the transition from the superfluid to 
the normal state occurring at the critical 
temperature $T_c$; and \textit{ii)} bosons at finite temperature, also near and 
above the Bose-Einstein condensation temperature $T_{BEC}$ 
($T_{BEC}$ coinciding with $T_c$ for Bose-Einstein condensates). 
The Boltzmann equation \cite{Landau10} 
provides a major tool to describe the collective 
dynamics of cold atoms in the normal-fluid phase. 
From such studies it emerged that the description 
with the Boltzmann equation 
(including the quantum statistics modification of the collision term \cite{Haug08}) 
works rather well not only for fermions above the Fermi temperature $T_F$ and for bosons above $T_{BEC}$, 
but also for two-components fermions at temperature smaller than $T_F$ and larger than $T_c$ 
\cite{Riedl2008,Chiacchiera2011,Pantel2012,Pantel2015}. 
We also mention that the dynamics of 
bosons at finite temperature below $T_{BEC}$ has been studied 
resorting to a Boltzmann-like description of the thermal part of the gas 
\cite{ZNG} -- a similar 
approach based on the Boltzmann description of the non-superfluid part of a two-component fermionic 
mixture has been discussed in \cite{Urban07}. Most of these 
studies have been performed in single-well potentials without tunneling between wells, also given 
the difficulty of computing in a quantitatively reliable way the relaxation time $\tau$ which is determined at finite temperature  
by rare events of hopping across the wells. 
Studies of the dynamics of a Bose gas below $T_{BEC}$ 
in a periodic multi-well potential 
were presented in \cite{Nikuni,Nikuni07,McK08}, 
while the study of the fermionic transport in optical lattices at finite 
temperature (above $T_c$) was reported in \cite{Rosch}, 
where the Boltzmann equation was investigated in 
local relaxation time approximation. 
The center of mass oscillations of a normal Fermi gas in a 
one-dimensional periodic potential were studied both 
theoretically and experimentally \cite{Pez04,Orso04}.

Although from one side a huge effort has been devoted to the study 
of quantum tunneling for superfluid atomic gases in double-well potentials 
at low temperatures, and from the other side the study of reaction-rate theory 
\cite{Hanggi90} and of the Boltzmann equation are two workhorses of 
non-equilibrium physics, the study of the Boltzmann equation itself 
in a double-well potential is to the best of our knowledge a relatively not 
addressed topic. 
Motivated by systems of cold atoms in tunable geometries at finite 
(and possibly) variable 
temperature, in this paper 
we therefore study the double-well dynamics of an interacting 
gas at finite temperature within the 
framework of the Boltzmann equation with the classical collision term.  

Our aim is to understand and describe, both qualitatively and quantitatively, the effect of two-body 
collisions on the the double-well dynamics. 
This problem is interesting in view of current and future experiments 
with cold atoms at finite temperature, since our approach can be applied 
to study the normal-fluid dynamics in double-well potentials.
Here we examine how a gas in a symmetric double-well, 
prepared with an initial population imbalance, 
relaxes towards the balanced equilibrium situation and we propose a simple 
analytical model to 
describe our numerical findings. 
Performing a comparison between numerical and analytical results,
we analyze the relaxation time as a function of barrier heights and interaction strengths.
We observe that, even though we are not going to consider 
the effect of the quantum statistics on the collision term, we expect that both the numerical computations and the 
analytical model can be straightforwardly extended 
to include such correction and that the relaxation time would 
have a similar dependence on 
barrier heights and interaction strengths, 
with qualitative changes of the dependence of $\tau$ 
on the temperature only for, say, $T/T_F \lesssim 0.5$ 
\cite{Riedl2008,Chiacchiera2011} 
(and therefore close to $T_c$ in the unitary limit). 

The article is organized as follows. In \Sec{sec:form} we introduce the 
formalism briefly describing the 
numerical method used in the work. We also 
define the two model external potentials we consider. 
The first model is a square-well potential with a barrier 
of vanishing width in the middle that provides a simplified 
toy model for our numerical and analytical study. 
The second model, which is directly inspired by experimental work with cold atoms, 
is the superposition of a harmonic isotropic potential and a gaussian barrier. 
In \Sec{sec:modelLowBarr} we introduce a simple 
analytical model for the study of the dynamics. 
In \Sec{sec:NumToy} and \Sec{sec:NumReal} we present 
the numerical results for, 
respectively, the square double-well potential 
and the realistic one and compare them with the predictions 
of the analytical model. 
Finally, we draw in \Sec{sec:concl} 
our conclusions and discuss possible improvements of the 
present paper for future work, while some useful results are collected 
in the Appendices.

\section{The Boltzmann equation with a double-well potential}
\label{sec:form}

In this Section we briefly recall the Boltzmann equation and 
introduce the two double-well potentials we consider in the following. 
Section \ref{sub-sec:Boltz_def} is devoted 
to introduce the Boltzmann equation formalism and remind 
some of its basic properties used hereafter, 
while in Section \ref{sub-sec:test} 
we briefly sketch the procedure 
for the numerical solution of the Boltzmann equation 
based on the test-particle method.  
Section \ref{sub-sec:models} is devoted to introduce the two double-well 
potentials studied in this paper: a ``toy'' 
square double-well (SDW) with a filtering 
wall and a more realistic harmonic-gaussian double-well (HGDW) potential. 
In Section \ref{sub-sec:times} we define the various time scales 
appearing in the double-well problem.
\subsection{Boltzmann equation}
\label{sub-sec:Boltz_def}

We consider a one-component gas of $N$ classical interacting particles 
of mass $m$, in an external potential $V(\vek{r})$.
The evolution in time of the phase-space distribution 
function $f(\vek{r},\vek{p},t)$ is governed by the Boltzmann equation 
\cite{Landau10}
\begin{equation}\label{BE}
\frac{\partial f}{\partial t} + \dot{\vek{r}} \cdot \nabla_{\vek{r}} f 
+ \dot{\vek{p}} \cdot \nabla_{\vek{p}} f = -I[f]~.
\end{equation}
The left-hand side, where $\dot{\vek{r}}={\vek{p}}/{m}$ and  $\dot{\vek{p}}=-\nabla_{\vek{r}} V$, 
represents transport. 
The right-hand side $I[f]$ is the collision integral
\begin{equation}
I[f] = \int \frac{d^3 p_1}{(2\pi\hbar)^3}  d\Omega \frac{d\sigma}{d\Omega} 
\frac{|\vek{p}-\vek{p}_1|}{m} \left( f f_1 - f' f_1'\right)~,
\end{equation}
where $\frac{d\sigma}{d\Omega}$ the differential cross section
and the notation $f,f_1,f^\prime,f^\prime_1$ is used as a shortcut 
for the distribution 
function evaluated at the same $\vek{r},t$, but with momenta 
$\vek{p},\vek{p}_1,\vek{p}^\prime,\vek{p}^\prime_1$, respectively. 
The normalization condition is $\int d\Gamma f = N$, 
and the average of any one-body 
variable $\mathcal{O}(\vek{r},\vek{p})$ is 
$\langle \mathcal{O} \rangle=\frac{1}{N}\int d\Gamma f \mathcal{O}$, 
where the phase-space 
volume element is $d\Gamma=d^3 r d^3 p /(2\pi\hbar)^3$.

At equilibrium, the distribution function reads
\begin{equation}
  f_{\eq}(\vek{r},\vek{p})=e^{-\beta\left(\frac{p^2}{2m}+V(\vek{r})-\mu\right)}~,
\end{equation}
where $\beta=1/(k_BT)$ is the inverse temperature and $\mu$ the chemical 
potential.
The total collision rate at equilibrium can be computed exactly using
\begin{equation}
\Gamma_\eq=\frac{1}{2}\int d^3r \int\frac{d^3p}{(2\pi\hbar)^3} \frac{d^3p_1}{(2\pi\hbar)^3}
d\Omega \frac{d\sigma}{d\Omega}\frac{|\vek{p}-\vek{p}_1|}{m}f_\eq f_{1,\eq}~,
\end{equation}
where the factor $1/2$ is needed to avoid double counting (since we 
are dealing with a one-component gas). Expressions for the collision rate 
in a single square-well and in a harmonic potential are collected 
in Appendix \ref{app:rates}.

In this work, we consider particles interacting with 
a constant cross section, i.e., with no energy or momentum dependence: 
\begin{equation}
\frac{d\sigma}{d\Omega}=\frac{\sigma}{4\pi}~.
\end{equation}
%

\subsection{Test-particle method}
\label{sub-sec:test}

A fully numerical approach to solve the Boltzmann equation 
is provided by the so-called {\it test-particle} method, in which the 
coordinates of all the particles in phase space are evolved individually.
This method, similar to molecular dynamics but with a stochastic component,
 was developed in the '80s 
in the context of nuclear physics \cite{Bertsch88} and recently 
has been applied also to cold atoms 
\cite{Jackson02,Toschi2003,Lepers2010,Wade2011,Goulko2011,Pantel2015}.
Formally, the distribution function $f(\vek{r},\vek{p},t)$ is discretized 
as a sum of delta-functions peaked at the position and momentum of each test particle:
\begin{equation}
f(\mathbf{r},\mathbf{p},t) = \frac{N}{\tilde N} \sum_{i=1}^{\tilde N} (2\pi \hbar)^3 
\delta\left(\mathbf{p} - \mathbf{p}_i(t)\right)
\delta\left(\mathbf{r} - \mathbf{r}_i(t)\right),
\end{equation}
where $N$ is the number of real particles and $\tilde N$ is the number of test particles, 
those entering the simulation. 
In some cases it is convenient to have 
$\tilde N \neq N$: if the number of real particles is very low, one 
associates to each real particle many 
test-particles ($\tilde{N}>N$), 
to describe the evolution in phase-space with a finer
resolution. At variance, in the opposite case of too many real particles 
to be simulated 
individually, one test particle represents many real ones 
($\tilde{N}<N$) \cite{Pantel2015}. We observe that 
for a generic $N/\tilde{N}$, the interactions between test-particles 
are ruled by a cross-section that is related to the real one as follows
\begin{equation}
\tilde\sigma = \frac{N}{\tilde N}\sigma~.
\end{equation}
In this work, we always 
take $N=\tilde{N}$.

The average value of any one-body 
observable $\mathcal{O}(\mathbf{r},\mathbf{p})$ is obtained as follows
\begin{equation}
\left<\mathcal{O}\right> = \frac{1}{N} \int d\Gamma  
f(\vek{r},\vek{p},t)\ \mathcal{O} (\mathbf{r},\mathbf{p})=
\frac{1}{\tilde N} \sum_{i=1}^{\tilde N} \mathcal{O}(\mathbf{r}_i,\mathbf{p}_i)~.
\end{equation}

In absence of inter-particle interactions, the phase-space coordinates 
of each test particle are evolved according to the Hamilton equations
\begin{equation}\label{eq:Ham}
\dot{\vek{r}}=\frac{\vek{p}}{m}\quad\quad\textrm{and}\quad\quad\dot{\vek{p}}=-\nabla_{\vek{r}} V(\vek{r})~.
\end{equation}
The actual numerical scheme to integrate them depends on the potential. In the case 
of a simple box, particles are propagated via the Euler method (exact in this case),
 and the collisions with the walls are implemented reversing the appropriate
 momentum component.
For a generic potential, the particle position and momentum are evolved 
from the time step $t_n$ to $t_{n+1}=t_n+\Delta t$ via 
the velocity Verlet algorithm \cite{Allen} 
\begin{equation}\label{eq:velVer}
\begin{array}{ll}
  \vek{v}(t_{n+\frac{1}{2}})=\vek{v}(t_n)+ \vek{a}(t_n)\Delta t/2\\
  \vek{r}(t_{n+1})=\vek{r}(t_n)+  \vek{v}(t_{n+\frac{1}{2}}) \Delta t\\
  \vek{v}(t_{n+1})=\vek{v}(t_{n+\frac{1}{2}})+ \vek{a}(t_{n+1})\Delta t/2
\end{array}~,
\end{equation}
with $\vek{a}(t)=-\nabla V(\vek{r}(t))/m$ the acceleration and 
$t_{n+\frac{1}{2}}$ an intermediate time step.

For interacting particles, collisions have to be implemented too.
The cross section defines an interaction range $d_\inter=\sqrt{\sigma/\pi}$: 
in each time step, all the pair of particles (within a certain distance) are 
checked, and they are collided if: {\it 1)} 
they reach their closest approach \cite{Lepers2010} in the time 
step; and {\it 2)} 
their distance at closest approach is within the interaction range. 
A collision takes place randomly assigning new momenta to the participants, 
with the constraints of energy and momentum conservation. 
In our simulations the trajectories of the 
colliding particles are corrected to take into account the fact that the 
collision takes place at the closest approach point \cite{Lepers2010}. 
For example, one can check that in the case 
of a harmonic well this correction is necessary to respect the balance of kinetic and potential energy. 

\subsection{Two models for the double-well}
\label{sub-sec:models}

We consider two model potentials for the double-well. The SDW potential 
has a rectangular energy barrier (of negligible width) 
located at the center of the system, which is in turn chosen to be a square 
well. This SDW has the advantage to be more simply numerically 
simulated using the test-particle method and it allows for an analytical 
treatment of the approximate model we are going to introduce 
in Section \ref{sec:modelLowBarr} 
for the determination of the relaxation times  
in double-well potentials. The other potential is  
relevant for cold-atom experiments and it is provided by the sum 
of a harmonic potential plus a barrier energy, chosen of gaussian 
form both for simplicity and also because such a barrier 
could be created by a superimposed 
blue-detuned potential. We refer to the second double-well potential 
as the harmonic-gaussian double-well (HGDW) potential.

The first model (SDW) is a toy model corresponding to a cubic box of side $2L$ 
partitioned into two regions by a filtering 
wall: i.e., particles are allowed to pass through it or are reflected 
depending upon their momentum component orthogonal to the wall (say, $p_x$).
When a particle during its propagation tries to cross the plane $x=0$, 
we check for its momentum along $x$: if $|p_x|>p_0$, where $p_0=\sqrt{2mV_0}$  
and $V_0$ the height of the barrier, then the particle it is allowed 
to pass, as if the barrier were not there; if instead
$|p_x|<p_0$, it is reflected by the barrier. 
This model can be seen as a finite barrier of negligible thickness 
$w \to 0$, and the corresponding potential reads
\begin{equation}\label{eq:VA1}
V(\vek{r})=\left\{
\begin{array}{ll}
V_0[\theta(x+\frac{w}{2})+\theta(\frac{w}{2}-x)-1]&, 
\,\, |x|,|y|,|z|\leq L\\  
\infty & ,\,\,\textrm{else}~.
\end{array}
\right.
\end{equation}
This model has the advantage to simplify both numerical and analytical 
approaches and it allows to study the effect of the barrier 
without introducing a specific shape for it.

The second potential (HGDW) is more realistic and it 
represents an isotropic harmonic potential (i.e., 
a spherical trap) plus a Gaussian barrier in one direction: 
\begin{equation}\label{eq:VB}
V(\vek{r})=\frac{m\omega_0^2r^2}{2}+ \tilde{V}e^{\frac{-x^2}{2w^2}}~.
\end{equation}
This is the structure of a realistic double-well potential in a cold-atom experiment.
\begin{figure}[t]
\begin{center}
\includegraphics[height=6cm,angle=0]{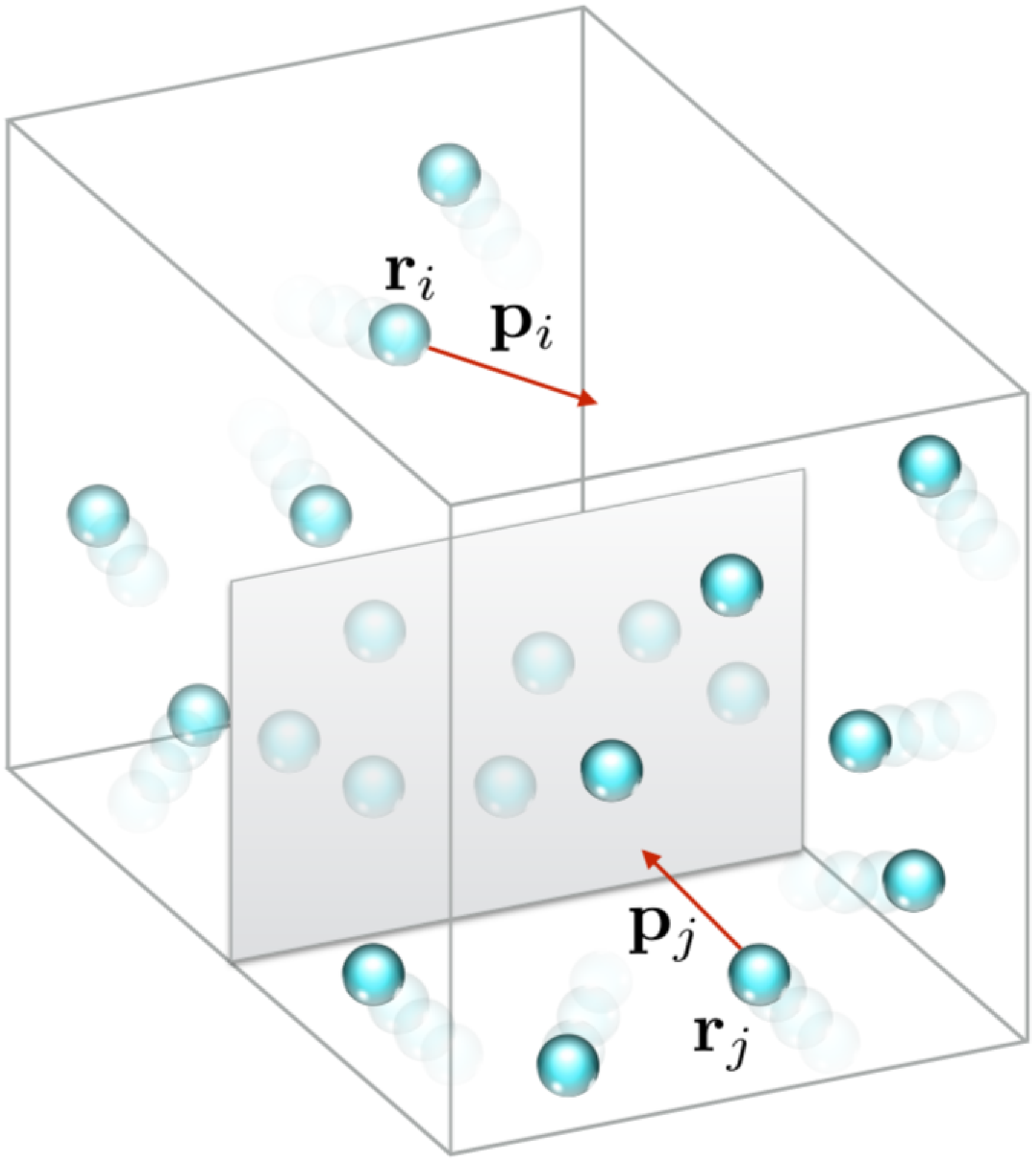}
\includegraphics[height=6cm,angle=0]{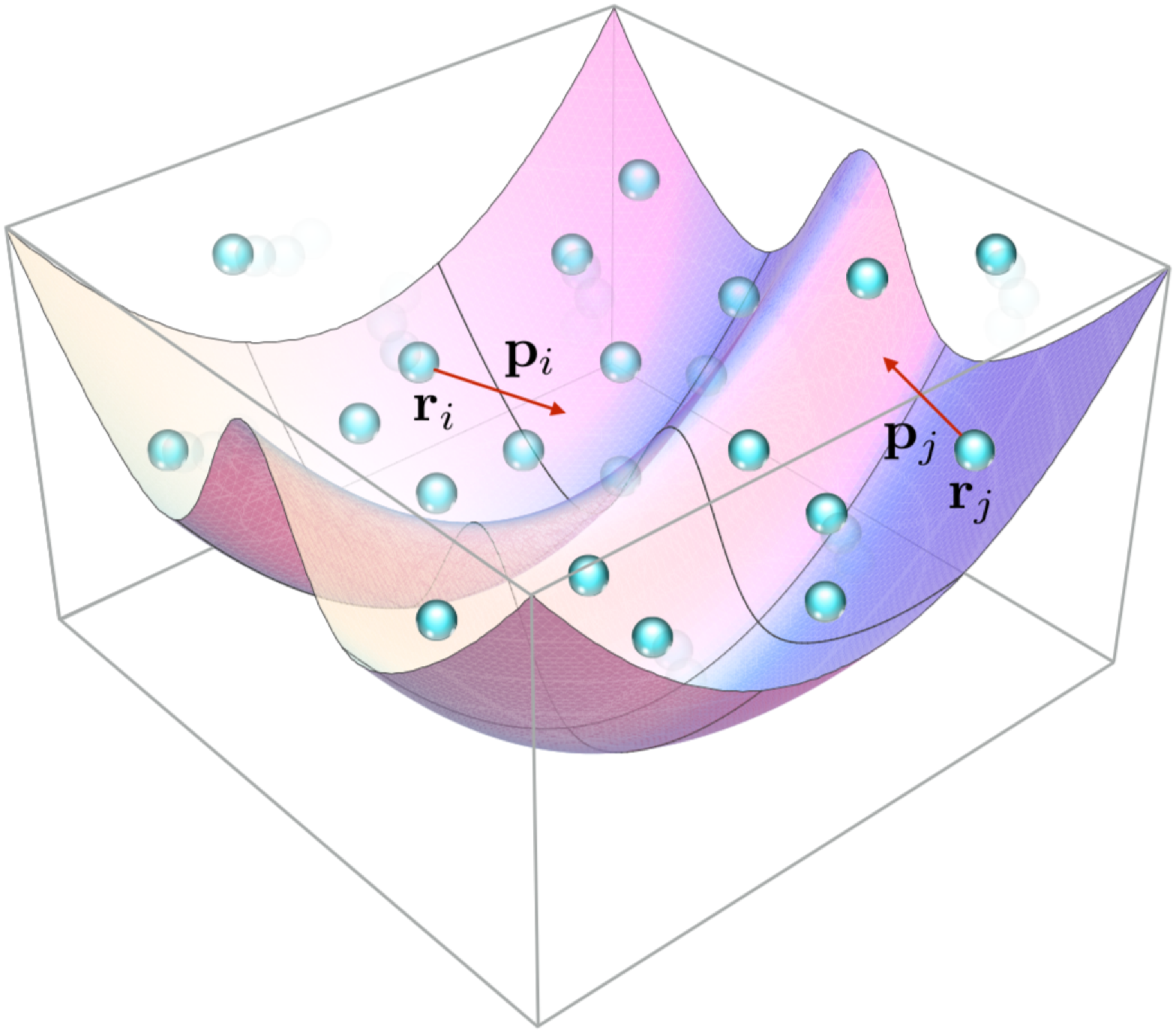}
\end{center}
\caption{Plot of the SDW (top) and HGDW (bottom) potentials. 
The figures represent $V(x,y,z=0)$, where $V$ is the potential energy. 
Particles are 
also pictorially shown, with their vertical coordinate representing 
their energy.}
\label{fig:wells}
\end{figure}

In \Fig{fig:wells} we represent schematically the two model potentials at $z=0$, 
as a function of $x$ and $y$. In this pictorial representation, the vertical 
coordinate of the particles corresponds to their energy.

In both cases, we denote as $V_0$ the barrier height: 
it is $V_0=p_0^2/2m$ for the SDW, and $V_0=V(\vek{0})-V(\vek{r}_{min})$, 
the difference of potential energy between the barrier top and the well 
minima in the case of the HGDW.

The quantity we use to follow the macroscopic dynamics of the double-well 
problem is the fraction of particles in the left well at time $t$
\begin{equation}
x_L(t)=\frac{N_L(t)}{N}~.
\end{equation}
The population imbalance is defined as 
\begin{equation}
\Delta N(t)=N_L(t)-N_R(t)~.
\end{equation}
In the literature, as in the case of the superfluid dynamics 
in double-well potentials, is also often used as well the 
relative imbalance $z(t)=\Delta N(t)/N$.

\subsection{Time scales}
\label{sub-sec:times}

There are in general 
two fundamental time scales that rule the dynamics of a trapped
interacting gas in a confining trap: one is related to the
inter-particle interactions, the other to the external potential. 

The average time $\tau_{\mathit coll}$ 
between two consecutive collisions experienced by 
the same particle is the collisional time
\begin{equation}\label{eq:taucoll}
\tau_{\mathit coll}= (2\Gamma_\eq/N)^{-1}~.
\end{equation}

On the other hand, the average time $\tau_\boxx$ 
a particle needs to travel across the 
whole trap is
\begin{equation}
\tau_\boxx= \frac{l_\boxx}{v_{\rms}}~,
\end{equation}
where the root-mean-square velocity at equilibrium is $v_\rms=\sqrt{3 /(\beta m)}$, 
and $l_\boxx$ the system linear size. 
The reference length $l_\boxx$ is chosen to be $2L$ for the SDW model (and 
as well for the single square-well treated in \Sec{sec:singletoy}) and 
$2\sqrt{2 \tilde{V}/(m\omega_0^2)}$ 
for the HGDW model (in the latter it is the distance between the 
two points of the harmonic trap having energy $\tilde{V}$, neglecting 
the barrier).

Depending upon the frequency of collisions, the gas can be in different 
dynamical regimes, the two limiting cases being the collision-less regime
(very rare collisions) and the hydrodynamical one (very frequent collisions):
if $\tau_\coll/\tau_\boxx\rightarrow \infty $ 
the gas is collision-less, whereas it is 
hydrodynamical if $\tau_\coll/\tau_\boxx\rightarrow 0$.

For the double-well problem, we are interested in the time 
needed to smooth out the initial population imbalance, and 
our aim is to relate it 
to the basic properties of the system.
This defines for double-well potentials a third time scale, 
related to the relaxation of imbalance, 
and that we will indicate with $\tau$.
Classically, the particles can cross the barrier only if their 
momentum perpendicular to it is high enough. 
Since the momentum is continuously redistributed 
between particles in collisional events, we expect that 
the relaxation time depends on $\tau_\coll$. 
Also, an Arrhenius-type of behaviour $1/\tau 
\propto \exp{(-\beta V_0)}$ is expected in the limit of large barriers 
($\beta V_0 \gtrsim 1$).

These three time scales ($\tau_{\mathit coll}$, $\tau_\boxx$, $\tau$) 
will of course depend on the potential 
considered. Throughout the next Sections, for different reasons, we 
deal with a square single-well, a square double-well, a harmonic oscillator
and a harmonic-gaussian double-well. To keep as light as we can the notation, 
we use the following convention: within any subsection, unless otherwise 
stated and denoted (respectively with ${\mathit sw}$, ${\mathit SDW}$, 
${\mathit HO}$ and ${\mathit HGDW}$),
we intend that the potential-dependent quantities are computed 
within the potential under consideration. 

\section{A simple analytical model}
\label{sec:modelLowBarr}

In this Section we develop a simple analytical model to describe the 
effect of collisions on the relaxation dynamics. 

This model takes inspiration from the tight-binding ansatz 
used for the Gross-Pitaevskii equation in double- or multi-well potentials 
\cite{Sme97,Jak97,Tro01}. 
The Gross-Pitaevskii equation, describing the dynamics 
of a superfluid, is an equation for a complex wavefunction $\psi$: 
one then introduces two degrees of freedom for $\psi$ (phase, $\varphi$, 
and number, $N$) and this leads for a double-well potential to the 
introduction of four degrees of freedom, two per well (say $\varphi_L$, 
$\varphi_R$ and $N_L$,$N_R$ in the left, $L$, and right, $R$, wells). 
These four degrees of freedom are not independent since the total number 
of particles $N_L+N_R$ is constant. Via these four non-independent 
degrees of freedom one then obtains a simplified, yet very good, description 
of the superfluid 
tunneling dynamics in the weakly coupled regime.

Of course, for a classical gas in a double-well potential, there is no 
tunneling and the distribution $f$ is not a complex number: anyway, 
one can still think 
to introduce suitable degrees of freedom per well and after 
coupling the two wells via the Boltzmann equation one obtains a description 
of the double-well dynamics. 
This leads to a simplified model, which allows for an 
approximate estimate of the relaxation time $\tau$. 
It is the choice of the degrees 
of freedom in the separate wells (effectively one in the model below, 
or eventually more for a more accurate 
description) that characterizes the model. Such choice 
is suggested by the form of the potential and by 
the properties of the physical system at hand: 
as an example, in \cite{DeWeert88} a simplified model 
was introduced to study the 
nonequilibrium distribution functions for electrons in the electrodes of a metal-insulator-metal junction.

In our case, we proceed to a rather drastic modelization of the Boltzmann 
equation dynamics taking into account that it is the energy barrier $V_0$ 
that suggests/sets an energy scale, dividing particles in two classes: the particles 
having energy larger than $V_0$ (which then can move from one well to the 
other) and the particles having energy smaller than $V_0$ (which do not). 
The mechanism for which particles can move from the latter class 
to the former is provided by interactions: two particles scatter below barrier 
and as consequence of the scattering one of the two, conserving energy, has 
an energy larger than $V_0$. 
Finer details, such as 
higher-order scattering process, are neglected. As 
a result the model is not expected to give a quantitatively 
accurate prediction of the relaxation time. Nevertheless, 
since the estimate of the relaxation 
time is in general a not simple problem, and what in particular is difficult 
is the determination of prefactors, it provides simple analytical formulas  
which are in general qualitatively reasonable. In 
the regime of intermediate barriers ($\beta V_0 \sim 1$) the agreement is 
found to be also quantitative. 
 
We start with the
single-well problem: this allows to classify the different types of 
collisions that will play a role in the double-well case and 
quantify their relative importance.

\subsection{Square single-well}\label{sec:singletoy}

Consider $N$ particles in a cubic box of volume $\Omega$ having size $2L$. 
This is the same potential that will be considered in the next 
\Sec{sec:doubletoy}, except the fact that there 
a filtering wall is added in $x=0$. 

In view of the double-well potential problem which we are interested in, 
we choose a reference momentum 
$p_0$ and a reference energy $V_0=p^2/2m$ (remember that here this choice 
is arbitrary since there is no barrier yet). We define $l_\boxx=2L$ and 
$N^>(t)$ [$N^<(t)$] the 
number of particles of the gas having $|p_x|>p_0$ ($|p_x|<p_0$) at time
 $t$. Clearly, $N^>(t)+N^<(t)=N$ at any time, and, at equilibrium, 
collisions maintain them to their equilibrium values $N^\gtrless_\eq$ 
(given in Appendix \ref{app:chem}).  
However, what is their evolution in time if the system starts 
from a situation with $N^\gtrless\neq N^\gtrless_\eq$? 

We make the following {\it assumption} 
for the distribution function:
\begin{equation}\label{eq:anssingle}
f(\vek{p},t) = g^>(\vek{p}) N^>(t) + g^<(\vek{p}) N^<(t)~,
\end{equation}
where
\begin{equation}
g^>(\vek{p}) \equiv e^{-\beta(\frac{p^2}{2m} -\mu^>)}\theta(p_x^2-p_0^2)
\end{equation}
and
\begin{equation}
g^<(\vek{p}) \equiv e^{-\beta(\frac{p^2}{2m} -\mu^<)}\theta(p_0^2-p_x^2)~.
\end{equation}
$\beta$ is a constant, and the chemical potentials $\mu^\gtrless$ are constant
(see Appendix \ref{app:chem}) 
and such that $\int d\Gamma ~g^\gtrless(\vek{p})=1$, so that 
$\int d\Gamma f=N^>(t)+N^<(t)=N$, as it should.
Notice that the choice \Eq{eq:anssingle} implies a distribution 
that is uniform in space, thermal for $p_y$ and $p_z$ and
has a discontinuity in the $p_x$ momentum distribution at $p_x^2=p_0^2$. 
Since the number of particles in the well is fixed there is only an independent 
parameter in \Eq{eq:anssingle} (given that the $g$'s are fixed by 
$\beta V_0$).

Let us now consider, among all the possible collisions, those that will 
alter $N^>$ and $N^<$. 
In a collision, each of the two incoming and outgoing particles can have 
momentum along $x$ above or below the reference values: 16 cases are therefore
possible. Among these, there are 6 type of processes altering $N^\gtrless$, 
namely:
\begin{eqnarray}
&&A: (<  <; < >), \quad B: (< >; < <) \nonumber\\
&&C: (> >; < >) ,  \quad D: (< > ; > >)\nonumber\\
&&E: (< < ; > >),  \quad  F: (> > ; < <)~. \label{ratedef} 
\end{eqnarray}
For example, with the notation $(< <; < >)$ we mean that in the collision 
the two incoming particles have $|p_x|< p_0$, whereas one of the outgoing 
ones (order does not matter) is above and one below reference. 
Let us indicate as $\Gamma_i$, $i=A,B,C,D,E,F$ 
the rates of each kind of process: 
the evolution in time of $N^\gtrless$ satisfies
\begin{equation}
\left\{
\begin{array}{l}
\dot{N}^>(t)= \phantom{-} \Gamma_A - \Gamma_B - \Gamma_C + \Gamma_D + 2 \Gamma_E- 2 \Gamma_F\label{eq:Nupdot}\\
\dot{N}^<(t)= -\Gamma_A + \Gamma_B + \Gamma_C - \Gamma_D - 2 \Gamma_E+ 2 \Gamma_F~,
\end{array}
\right.
\end{equation}
and $\dot{N}^>(t)+\dot{N}^<(t)=0$, as it should.
Each term $\Gamma_i$ is multiplied by an appropriate factor 
taking into account the changes in $N^\gtrless$ the process implies. 
By a direct computation of the rates it is found that
\begin{eqnarray}
& \Gamma_A= \gamma_1 (r^<)^2; \quad  &\Gamma_B= \gamma_1 r^> r^< \nonumber\\
& \Gamma_C= \gamma_2 (r^>)^2; \quad  &\Gamma_D= \gamma_2 r^> r^< \nonumber\\
& 2\Gamma_E= \gamma_3 (r^<)^2; \quad &2\Gamma_F= \gamma_3 (r^>)^2 \label{eq:Gammas}
\end{eqnarray}
where $r^\gtrless(t)\equiv N^\gtrless(t) /N^\gtrless_\eq$. 
At equilibrium, $r^\gtrless_\eq=1 $ and the rates of the processes 
just defined are equal two by two: $\Gamma_A^\eq=\Gamma_B^\eq$, 
$\Gamma_C^\eq=\Gamma_D^\eq$, and $\Gamma_E^\eq=\Gamma_F^\eq$.
The factors 
$\gamma_i$, $i=1,2,3$ are appropriate equilibrium phase space integrals 
detailed in Appendix \ref{sec:h1h2h3}. They are proportional 
to the equilibrium collision rate: in fact, they are given by 
\begin{equation}\label{eq:gammah}
\gamma_i=\Gamma_\eq h_i(\beta V_0),\quad i=1,2,3~,
\end{equation}
where $V_0=p_0^2/2m$ the reference energy and 
$h_i$ are adimensional functions (see Appendix \ref{sec:h1h2h3}).

Inserting these results into \Eq{eq:Nupdot} we get
\begin{equation}\label{eq:eomNup}
\dot{N}^> = -\Gamma_\eq(r^>-r^<)[(h_1 + h_3) r^< + (h_2 + h_3) r^> ]~,
\end{equation}
or, equivalently, dividing by $N$:
\begin{equation}\label{eq:eomxup}
\dot{x}^>\equiv\frac{\dot{N}^>}{N} = -\frac{\Gamma_\eq}{N}(r^>-r^<)[(h_1 + h_3) r^< + (h_2 + h_3) r^> ]~.
\end{equation}
Therefore, according to this model, the relaxation dynamics of the single-well 
problem depends only upon the parameter $\beta V_0$; 
all the others ($T$, $N$, $\sigma$, $m$, and the volume $\Omega$) 
enter only through the combination $N/\Gamma_\eq$ 
[see Appendix \ref{app:rates}, \Eq{eq:Gammaeqbox}] 
with the result of setting a time scale for the 
problem, i.e. a prefactor entering the time $\tau$ in which 
the particles move from above (below) to below (above) the reference 
energy.

The stationarity condition is 
\begin{equation}
r^< = r^> 
\Leftrightarrow  N^>=N^>_\eq,  N^<=N^<_\eq~.
\end{equation}

The fraction of particles that at equilibrium is above reference 
(have $p_x^2>p_0^2$) can be easily evaluated and is
\begin{equation}\label{eq:xaboeq}
x^>_\eq\equiv\frac{N^>_\eq}{N}= \erfcB~.
\end{equation}

\begin{figure}[t]
\begin{center}
\includegraphics[width=8.cm,angle=0]{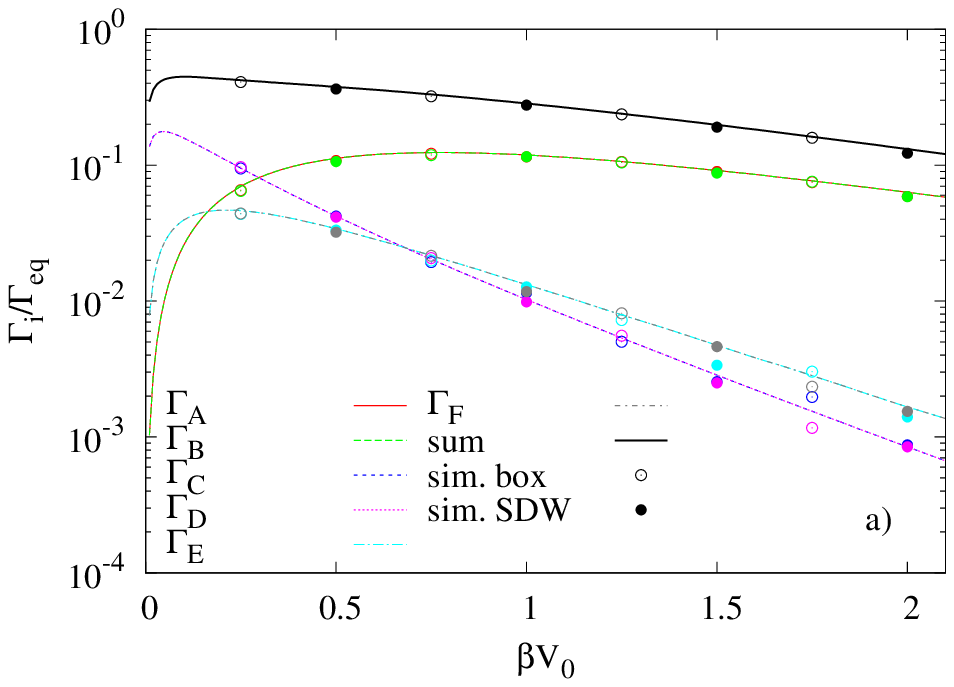} 
\includegraphics[width=8.cm,angle=0]{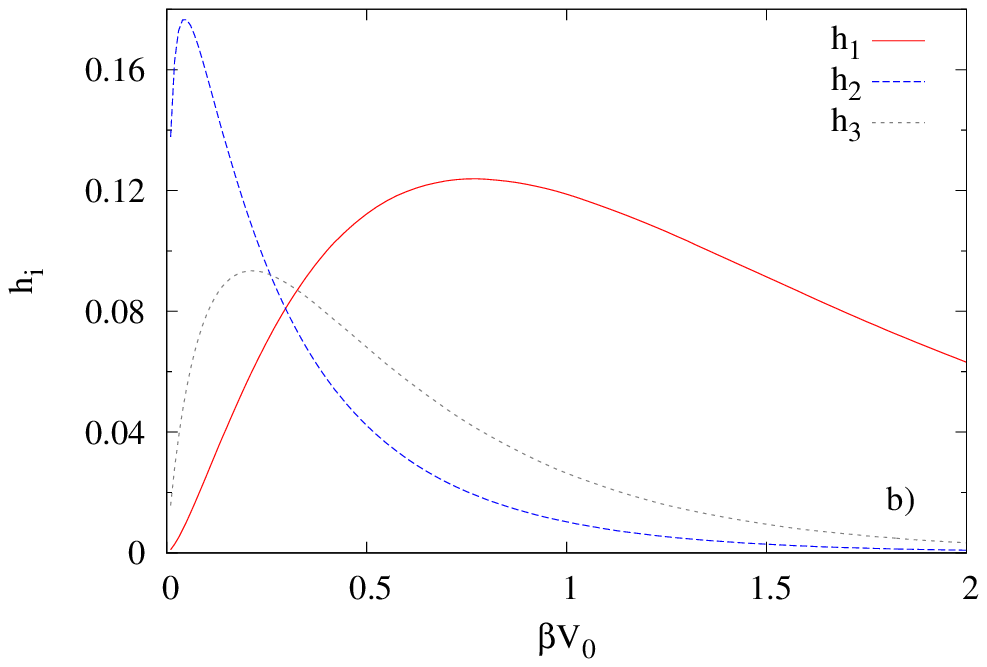} 
\end{center}
\caption{Rates at equilibrium. 
Panel {\it a)}: equilibrium rates of the processes $A$, $B$, $C$, $D$, $E$, and 
$F$, defined in (\ref{ratedef}), 
and their sum {\it vs} $\beta V_0$. 
The rates are equal two by two and correspond to $h_1,h_2,h_3/2$. 
The simulations (points) were performed for a square single-well potential, 
with parameters $\tilde{N}=5000$, $L=0.5$, $T=10$. 
Results with (full circles) and without (open circles) a barrier are shown.  
The theory prediction is shown for comparison (lines). 
Panel {\it b)}: 
the adimensional functions $h_1$, $h_2$ and $h_3$ (with linear scale in 
vertical axis) {\it vs} $\beta V_0$. 
}
\label{fig:rates}
\end{figure}

In panel {\it a)} of \Fig{fig:rates}
we show (lines) the various rates at equilibrium {\it vs} 
$\beta V_0$. The rates 
obtained in the simulation at equilibrium are also shown (points) as a check of the numerical 
algorithm. The empty circles denote the rates in a single-well, and 
the full circles denote the rates in a double-well. 
The presence of the filtering wall prevents collisions between particles
 on different sides, unless both have high enough energy; 
we also verified that $\Gamma_i$ and $\Gamma_\eq$ are separately 
very similar for the single square-well and the SDW since 
the effect of the tiny barrier on these  
bulk quantities is negligible.

From panel {\it a)} of \Fig{fig:rates} we also see that,
 out of all the collisions taking place, 
the processes affecting $N^\gtrless$ (full line) are, on the whole, 
always a percentage smaller than $50\%$ and, for $\beta V_0=2$ have 
already decreased to $\sim 10\%$. In our simulations, we will always 
consider $\beta V_0<2$. Concerning the relative importance 
of the different processes, $A$ and $B$ are the most probable ones 
for $\beta V_0\gtrsim 0.5$.

For completeness, we show in panel {\it b)} the adimensional functions
 $h_i(\beta V_0)$, $i=1,2,3$; notice that these quantities, strictly related 
to those shown in panel {\it a)} [see \Eq{eq:Gammas} and \Eq{eq:gammah}], 
are shown here using a linear scale in the vertical axis. 

Finally, we would like to emphasize that, while of course the total collision 
rate $\Gamma_\eq$ depends on the trap shape 
(see Appendix \ref{app:rates}), the ratios 
$\Gamma_i/\Gamma_\eq$ do not (see Appendix \ref{sec:h1h2h3}). In fact, 
the spatial dependence factors out and cancels 
between numerator and denominator, 
leaving only integrals over the momenta. 
Therefore, also the data obtained in a realistic 
trap would fall on top of the curves of \Fig{fig:rates}a.

We now discuss the linearization of the dynamical equations 
\Eq{eq:eomNup} and the drawbacks of the analytical model.

\subsubsection{Linearization and relaxation time}

The evolution in time of $N^>$ is ruled by the non-linear equation 
\Eq{eq:eomNup}. If linearized around the equilibrium, it admits an 
exponentially decaying solution $N^>(t)=N^>_\eq+b~e^{-t/\tau}$ with 
the single-well relaxation time $\tau$ that (omitting details) reads 
\begin{equation}\label{eq:tausw}
\tau=\frac{N}{\Gamma_\eq}\frac{\erfcB~\erfB}{h_1+h_2+2h_3}
= 2 \tau_\coll~ \frac{x^>_\eq(1-x^>_\eq)}{h_1+h_2+2h_3}~.
\end{equation}
In the last equality we used the relations \Eq{eq:taucoll} 
and \Eq{eq:xaboeq}. 

\subsubsection{Shortcomings of the analytical model}

The simple assumption 
\Eq{eq:anssingle} is expressed in terms of just 
one dynamical degree of freedom, $N^>(t)$: 
this choice does not allow to respect 
energy conservation exactly. The energy of our gas is purely kinetic and 
can be computed at any time from the distribution function as
\begin{equation}
E(t)=\int d\Gamma f(\vek{p},t) \frac{p^2}{2m}~.
\end{equation}
At equilibrium, it yields correctly 
$E_\eq=\frac{3}{2}\frac{N}{\beta}$. However, 
the assumption 
\Eq{eq:anssingle} implies the following energy variation
\begin{equation}
\frac{\delta E(t)}{E_\eq}=\frac{N^>(t)-N^>_\eq}{N}~C(\sqrt{\beta V_0})~,
\end{equation}
where 
\begin{equation}
C(x)=\frac{2}{3\sqrt{\pi}}~
\frac{x~e^{-x^2}}{\erfc(x)~\erf(x)}~.
\end{equation}
The function $C$ is positive for $x\geq 0$: 
it starts from $C(0)=1/3$ and increases 
for larger values of $x$.
So, by construction, energy conservation is violated in the analytical model 
as soon as $N^>\neq N^>_\eq$. 
We can expect the model to be better 
as long as $|\delta E(t) /E_\eq|$ remains 
small enough during the 
whole evolution. However, we explored different initial conditions 
having the same $\Delta N (t=0)$ and we observed that even in cases 
in which $|\delta E(t) /E_\eq|$ become relatively large, the estimate of 
$\tau$ is good due to a compensation of effects. It also emerged that 
the estimate of $\tau$ appears to be better when $\delta E$ is positive.

The limitations due to the non-conservation of the energy 
could be overcome by choosing 
a structure for $f(\vek{p},t)$
with at least a further independent degree of freedom: for example, 
one could introduce a variable giving the amount of energy above reference 
energy $V_0$.

\subsection{Square double-well}\label{sec:doubletoy}

We now insert an energy barrier inside the box of \Sec{sec:singletoy}, 
obtaining the SDW potential. 
To be specific, assume the box is $[-L,L]^3$ and the barrier located 
at $x=0$. The barrier is perfectly transparent for particles with momentum 
$|p_x|>p_0$ and perfectly reflecting for particles with momentum $|p_x|<p_0$. 
The energy associated to the barrier is $V_0=p_0^2/2m$.

We can now straightforwardly extend the model seen for the single-well 
to this case: we write 
\begin{equation}
f(\vek{p},t) = f_L(\vek{p},t)+f_R(\vek{p},t),
\end{equation}
as in \Eq{eq:anssingle} with 
$f_\alpha(\vek{p},t) = g_\alpha^>(\vek{p}) N_\alpha^>(t) + 
g_\alpha^<(\vek{p}) N_\alpha^<(t)$ and $\alpha=L,R$ denoting the well. 
Since the role played by $V_0$ is the same as in the single-well model, 
the $g$'s depends on $\beta V_0$ in the same way and they do not 
depend on $\alpha$: $g_L=g_R\equiv g$. 
The numbers of particles in the two wells are $N_L=N_L^>+N_L^<$ and 
$N_R=N_R^>+N_R^<$.
Now we have four variables/components 
$N_L^>$, $N_L^<$, $N_R^>$, $N_R^<$. The equations of 
motion are 
\begin{widetext}
\begin{equation}
\left\{
\begin{array}{l}
\dot{N}^>_L = -\Gamma_\eq^\prime(r^>_L-r^<_L)[(h_1 + h_3) r^<_L + (h_2 + h_3) r^>_L ]- k_A(N^>_L-N^>_R)\\
\dot{N}^<_L = \phantom{-}\Gamma_\eq^\prime(r^>_L-r^<_L)[(h_1 + h_3) r^<_L + (h_2 + h_3) r^>_L ]\\
\dot{N}^>_R = -\Gamma_\eq^\prime(r^>_R-r^<_R)[(h_1 + h_3) r^<_R + (h_2 + h_3) r^>_R ]+ k_A(N^>_L-N^>_R)\\
\dot{N}^<_R = \phantom{-}\Gamma_\eq^\prime(r^>_R-r^<_R)[(h_1 + h_3) r^<_R + (h_2 + h_3) r^>_R ]~.
\end{array}
\right.
\label{eq:dw}
\end{equation}
\end{widetext}
We have denoted with a $^\prime$ the quantities referring to $N^\prime\equiv N/2$
 particles in a volume  $\Omega^\prime\equiv \Omega/2$: of course the density is the 
same and therefore the collision rate per particle is the same too 
($\Gamma_\eq^\prime/N^\prime=\Gamma_\eq/N$). 

We observe that the numbers 
$N_L^>$, $N_L^<$, $N_R^>$, $N_R^<$ obey 
the condition $N_L^>+N_L^<+N_R^>+N_R^<=N$ ($N$ is the total number of particles 
in the double-well) 
resulting in {\it three} independent parameters. 
However, during the first part of the dynamics, the particles above 
the barrier rapidly flow from one well to the other practically giving 
$N_L^> \approx N_R^>$, and therefore the independent parameters in the 
subsequent dynamics are just two.

With respect to the single-well case \Eq{eq:eomxup},
there is a qualitatively new term coupling the $L$ and $R$ sides of the 
barrier: it is a diffusion term giving the rate of particles passing from one side of the wall to the other
\begin{equation}
\dot{N}^>_L\Big\vert_\diff= - \frac{\Delta N_{L\rightarrow R}}{\Delta t} +
\frac{\Delta N_{R\rightarrow L}}{\Delta t} = -k_A (N^>_L- N^>_R)~.
\end{equation}
The coefficient $k_A$ (Arrhenius) is
\begin{equation}
k_A=  \frac{1}{L\sqrt{2\pi\beta m}}~\frac{e^{-\beta V_0}}{\erfcB}
~.
\end{equation}

A global factor $1/L$ in $k_A$ arises from the ratio between the area of the filtering 
wall and the volume of each partition: it would be replaced, in a more general 
geometry, by $\Sigma/\Omega^\prime$.

\subsubsection{Linearization and relaxation time}

Linearizing \Eqs{eq:dw} around equilibrium, we find they admit 
the following solution
\begin{equation}
x_L(t)=\frac{1}{2}+c_1~e^{\lambda_1 t}+c_2~e^{\lambda_2 t}~,
\end{equation}
where the eigenvalues are found to be 
\begin{eqnarray}
\lambda_{1,2} &=& -\frac{1}{2} \left(2k_A+ k_{\mathit sw} 
\pm \sqrt{\left( 2k_A+ k_{\mathit sw} \right)^2 -8 k_{\mathit sw}
 k_A x^>_\eq}  \right)\nonumber\\
&\equiv&-\frac{1}{\tau_{1,2}}~.
\end{eqnarray}
We have defined the single-well rate $k_{\mathit sw} 
\equiv\frac{1}{\tau_{\mathit sw}}$, with $\tau_{\mathit sw}$ the single-well 
relaxation time given by \Eq{eq:tausw}. 
We also denoted the eigenvalues so that 
$\tau_1$ is larger than $\tau_2$.
In the comparison with the numerical results 
(see next sections), since $\tau_2$ is associated to the short-time 
dynamics and $\tau_1$ to the long-time one, we plot $\tau_1$ as 
the analytical prediction for the relaxation time. 
In the limit of large cross sections, $\tau_1$ tends to 
the diffusive limit
$\tau_{1,\diff}\equiv\lim_{\sigma\rightarrow \infty} \tau_1 =
L \sqrt{\pi\beta m/2}~e^{\beta V_0}$. 

Since $x^>_\eq$ is small, one can get more insight approximating 
the eigenvalues 
to lowest order in $x^>_\eq$:
\begin{equation}
\tau_1\simeq \left(\frac{1}{k_\sw}+\frac{1}{2k_A} \right)\frac{1}{x^>_\eq}~,
\quad \quad
\tau_2\simeq \frac{1}{2k_A+k_\sw}~.
\end{equation}
Notice that $\tau_1$ is a sum of two terms, one depending 
on the collision (to which we may refer as 
$\tau_{1,\coll}$) and the other not ($\tau_{1,\diff}$). The latter term 
is dominating for large $\sigma$, and for all $\sigma$ 
it is $\tau_1 \ge \tau_{1,\diff}$.

We can also write $\tau_1$ as 
\begin{equation}
\tau_1 \simeq L \sqrt{\beta m}\left[
\frac{4}{\sqrt{\pi}} \frac{1}{N\left(\frac{d_\inter}{L}\right)^2}
\frac{\erfB}{h_1+h_2+2h_3} + \sqrt{\frac{\pi}{2}} e^{\beta V_0}
 \right]
\end{equation}
where $d_\inter=\sqrt{\sigma/\pi}$  is the length scale associated to
interactions. 

Summarizing, in the cubic box with a filtering barrier 
(the DSW model) we get:
\begin{equation}
-\frac{1}{\tau_1}=-\frac{1}{2} \left(2k_A+ k_\sw
- \sqrt{\left( 2k_A+ k_\sw \right)^2 -8 k_\sw
 k_A x^>_\eq}  \right)~,
\label{eq:tau1}
\end{equation}
where the coefficients are 
\begin{eqnarray}
&k_\sw&=\frac{\Gamma_\eq}{N}\frac{h_1+h_2+2h_3}{x^>_\eq(1-x^>_\eq)}\label{eq:kstoy1}\\
&k_A&=\frac{1}{L\sqrt{2\pi\beta m}}\frac{e^{-\beta V_0}}{x^>_\eq}~,\label{eq:kstoy2}
\end{eqnarray}
and $x^>_\eq$ is given in \Eq{eq:xaboeq}. 

\subsection{Adapting the analytical model to the 
harmonic-gaussian double-well}\label{sec:adapt}

As an approximation for the realistic double-well, we suppose we can still use
\Eq{eq:tau1}, but with other coefficients:
\begin{eqnarray}
&k_{\sw}^{\mathit HGDW}&=\frac{2\Gamma_{\eq,\textrm{HO at min},N/2}}{N}\frac{h_1+h_2+2h_3}{x^>_\eq(1-x^>_\eq)}\label{eq:ksreal1}\\
&k_{A}^{\mathit HGDW}&=\frac{\omega_\textrm{min}}{2\pi}\frac{e^{-\beta V_0}}{x^>_\eq}~,\label{eq:ksreal2}
\end{eqnarray}
with $x^>_\eq$ as given in \Eq{eq:xaboeq}.

In fact, the rates $h_i$ are the same in any potential $V({\bf r})$, 
because they are global quantities in which spatial dependence cancels 
out (see Appendix \Sec{sec:h1h2h3}). 
What changes when passing from the toy to the realistic double-well are the 
rate of collisions, entering $k_\sw$, and the characteristic frequency, 
entering $k_A$. In the Eqs. (\ref{eq:ksreal1}) and (\ref{eq:ksreal2}) we have used the harmonic approximation 
of the realistic HGDW well close to its minima, where 
$\omega_\textrm{min}^2=2\log{(\tilde{V}/m\omega_0^2w^2)}$ and 
$\Gamma_{\eq,\textrm{HO at min},N/2}$
 is the rate for $N/2$ particles in a harmonic trap of frequencies 
$(\omega_x,\omega_y,\omega_z)=(\omega_\textrm{min},\omega_0,\omega_0)$ 
(see Appendix \ref{app:rates}).
As a consequence of the modification of $k_A$, also the expression of the 
diffusive limit of $\tau_1$ is altered into 
$\tau_{1,\diff}^{\mathit HGDW}= \frac{\pi}{\omega_\textrm{min}}~e^{\beta V_0}$.

\section{Numerical results for the square double-well}\label{sec:NumToy}

We show in this Section the numerical results of 
test-particle simulations 
for the SDW model potential. We consider a system with given 
$N$, density and temperature ($N=5000$, $L=0.5$, $T=10$) and vary 
the barrier height $V_0$ and the cross section $\sigma$. 
In all the simulations, the initial 
population is $60\%$ in the left well and $40\%$ in the right one.
A comment is in order about units: in this Section we use units in 
which $\hbar=k_B=m=1$ and $2L=1$.

\begin{figure}[t]
\begin{center}
\includegraphics[width=8cm,angle=0]{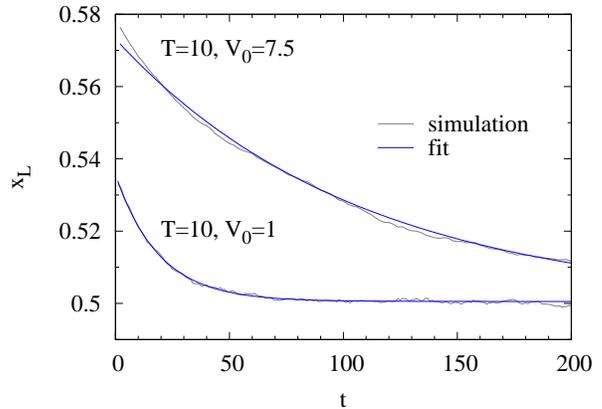}
\end{center}
\caption{Dynamics in the SDW potential. Time evolution of the well population 
for two barrier heights. 
Parameters: $N=5000$, $L=0.5$, $T=10$, $\sigma=\pi \times(0.001)^2$. 
The barrier height is $V_0=1$ for the lower curve and 
$V_0=7.5$ for the upper one. The simulation curves are obtained averaging 
over $40$ runs with different microscopic initial conditions and smoothing over 
small time intervals. 
The fits are done with the function $f_{1}(t)=a+b\,e^{-t/\tau}$.
}
\label{fig:xL-toy}
\end{figure}

As an example, we show in \Fig{fig:xL-toy}
the evolution in time of population imbalance for two
values of $V_0$. The curves are obtained averaging over $40$ runs 
with different microscopic initial conditions and smoothing over small time 
intervals. These two procedures are necessary because the barrier crossing 
is a rare event and leads to large fluctuations in the well population.
Notice that this is not a numerical artifact: also in an experimental 
realization it would be necessary to average over different realizations to be able to observe 
the population evolution in time clearly for large barriers.
As a consequence of the smoothing, the curves in \Fig{fig:xL-toy} start from 
values different from $0.6$: for example, 
in the case of a very low barrier, $V_0=1$, 
the adjustment from $x_L=0.6$ to $x_L\simeq0.53$ was very fast.

The obtained behaviour can be nicely fitted with a single exponential decay 
\begin{equation}
f_{1}(t)=a+b\,e^{-t/\tau}
\end{equation}
that allows to extract the relaxation time $\tau$. 
For other choices of the cross section or other barrier heights, 
the evolution in time of $x_L(t)$ is sometimes more complex, showing an 
oscillatory behavior 
at the early times or an exponential decay with two clearly separated 
time scales. 
Therefore, in the following, to extract the relaxation time $\tau$ from $x_L(t)$, 
we will also consider the fitting functions 
$f_2(t)=a+b\,e^{-t/\tau} +c \cos(\omega t+d) e^{-t/\tau_2}$ and 
$f_3(t)=a+b\,e^{-t/\tau} +c\,e^{-t/\tau_2}$.

\begin{figure}[t]
\begin{center}
\includegraphics[width=6.5cm,angle=0]{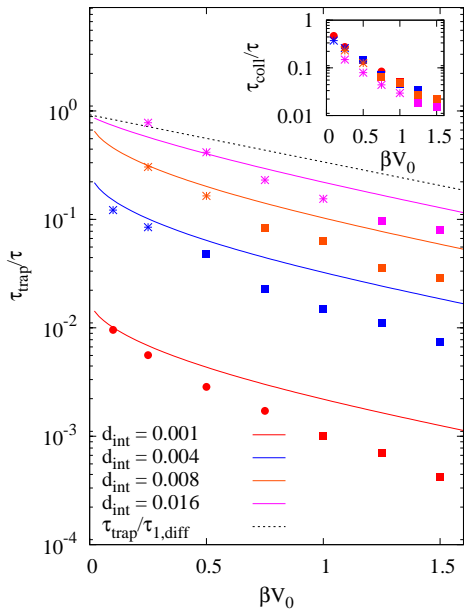} 
\end{center}
\caption{SDW potential relaxation times. The inverse relaxation time $1/\tau$, 
in units of $1/\tau_\boxx$,
as a function of $\beta V_0$ for different values of the cross section:
$\sigma=\pi \times(0.001, 0.004, 0.008, 0.016)^2$ 
with $x_L(t=0)=0.6$. The numerical results 
are denoted by points. 
The system is a box of size $2L=1$, with $N=5000$ particles, at $T=10$ and 
different barriers $V_0$. 
Each point is obtained from an average over $40$ runs 
($64$ for the largest value of the cross section 
$\sigma=\pi \times 0.016^2$) and smoothing over small time intervals. 
The fits are done in the whole time interval, 
using one of the three functions 
$f_1$ (full circles), $f_2(t)$ (stars) and 
$f_3(t)$ (full squares) defined in the text. 
Different colors indicate different values of the cross section.
For comparison, we show (solid lines) the results obtained in an approximated 
analytical solution presented in \Sec{sec:modelLowBarr}, 
namely $\tau_1$ obtained 
using Eqs. (\ref{eq:tau1}), (\ref{eq:xaboeq}), (\ref{eq:kstoy1}) and (\ref{eq:kstoy2}).
The diffusive limit $\tau_{1,\diff}$ is also shown (dashed line).
Inset: $\tau$ rescaled in units of collisional time $\tau_\coll$.  
Colors correspond to the same cross section as the main figure.
}
\label{fig:tauVSbetaV0}
\end{figure}

In \Fig{fig:tauVSbetaV0}
we show the results for the relaxation time for an initial imbalance of $20\%$ 
[$x_L(t=0)=0.6$], and different values of the cross section 
$\sigma=\pi\times(0.001,0.004,0.008,0.016)^2$ as a function of $\beta V_0$.
The relaxation time is shown in units of $\tau_\boxx$, that for the 
toy model it $\tau_{\boxx}=2L\sqrt{\beta m/3}$. Any point 
is the result of a fit done on a curve 
obtained averaging $40$ ($64$ in the case of the largest cross section) 
runs with different 
microscopic initializations. The fits are done on the whole available time 
interval, and the 
chosen fitting function is the one leading to the smaller $\chi^2$. 
The point colour (on-line) indicates the value of the cross section, 
the point shape the fitting function: $f_1$ (full circles), $f_2$ (stars), 
$f_3$ (full squares). 
In the inset we show again the inverse relaxation time, but in units 
of the collisional time: 
in this scale all the points fall in a narrow region 
(notice that the range of values in the 
vertical axis of the inset is much smaller than in the main plot), 
showing that it is the collisional 
time that gives the major contribution to $\tau$.

\begin{figure}[t]
\begin{center}
\includegraphics[width=8cm,angle=0]{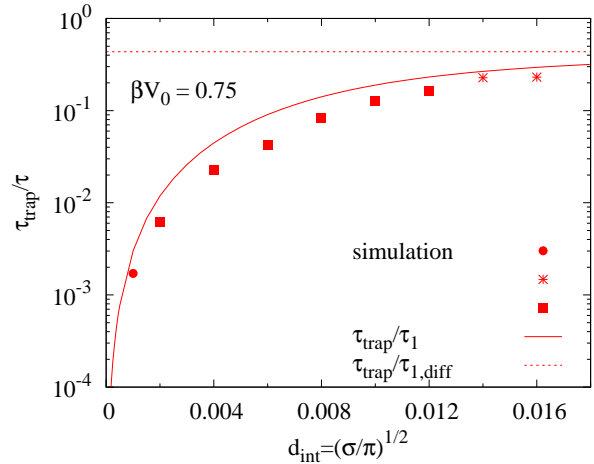} 
\end{center}
\caption{SDW model potential. Inverse relaxation time $1/\tau$, 
in units of $1/\tau_\boxx$,
as a function of 
$d_\inter$ for a given barrier height, $\beta V_0=0.75$. Parameters 
and the meaning of symbols and lines are the 
same of \Fig{fig:tauVSbetaV0}. 
The numerical results are denoted by points and 
the model prediction by lines.
}
\label{fig:tauVSd}
\end{figure}

In \Fig{fig:tauVSd}
we focus on a single barrier height and show the dependence of $\tau$ 
(again in units of $\tau_\boxx$) upon the interaction strength.

\subsubsection{Comparison with the analytical model predictions}

In \Fig{fig:tauVSbetaV0} and \Fig{fig:tauVSd} we compare the numerical 
results with the predictions of the analytical model for the SDW (lines). 
The model in general  
predicts a faster relaxation (smaller $\tau$) than what found in the 
simulation, and as the barrier height increases the agreement is deteriorated. 
In \Fig{fig:tauVSbetaV0} one sees, however, that for low and intermediate 
barriers the agreement is rather satisfactory.

A feature that the model captures nicely is the dependence of the relaxation time upon the interaction strength: 
in \Fig{fig:tauVSbetaV0} one can see that, as the cross section increases, 
both types of curves accumulate towards the diffusive limit.
This finding is better seen in \Fig{fig:tauVSd}, where we fix the
 relative barrier height at an intermediate value ($\beta V_0=0.75$) 
and study the evolution of $\tau$ 
with the interaction strength.

\section{Numerical results for the harmonic-gaussian double-well}
\label{sec:NumReal}

In this Section we present our results for the realistic HGDW, 
\Eq{eq:VB}, that is a combination of a spherical harmonic trap and a gaussian 
barrier along one direction. 
We start with an initial configuration with $60\%$ of the particles in the left 
well and $40\%$ in the right one; their initial momentum distribution is the 
thermal equilibrium one.
To prepare this configuration numerically, we start with the balanced 
population and then move $10\%$ of the particles from the right to the left well.

We then let the system evolve: $x_L(t)$ results 
from a combination of a center of mass oscillation due to the harmonic trapping, 
and the relaxation of population imbalance. We choose to fit such behaviour 
with the $f_2$ fit function 
$f_2(t)=a+b\,e^{-t/\tau} +c \cos(\omega t+d) e^{-t/\tau_2}$, where 
the third term represents the damped center of mass oscillation. 
We have checked this interpretation exciting explicitly the 
c.o.m. motion, i.e., considering a balanced cloud and 
displacing it on the whole by a certain amount. The frequency and damping 
of this c.o.m. mode are in reasonable agreement 
with those extracted from the oscillation 
that arises when the system is prepared with an initial population imbalance and 
not displaced. 

We fix the double-well shape (i.e., $\omega_0$, $w$, $\tilde{V}$) 
and study the 
relaxation dynamics for different temperatures and interactions. 
In all the cases 
we have $N=5000$ particles and the initial imbalance is $20\%$. 
The curves are obtained averaging over $40$ different 
microscopic realizations and smoothing over small time intervals. 
Since we are in presence of a harmonic trap, 
in this Section we use the harmonic oscillator units (or trap units), 
in which all the 
dimensional quantities are built from $\hbar,k_B,m,\omega_0$ as usual. 
For example, $E_\ho=\hbar\omega_0$, $l_\ho=\sqrt{\hbar/m\omega_0}$ and so on. 
To pass to physical units, values have to be chosen for the mass $m$ and the 
trap frequency $\omega_0$.

\begin{figure*}[t]
\begin{center}
\includegraphics[width=5.9cm,angle=0]{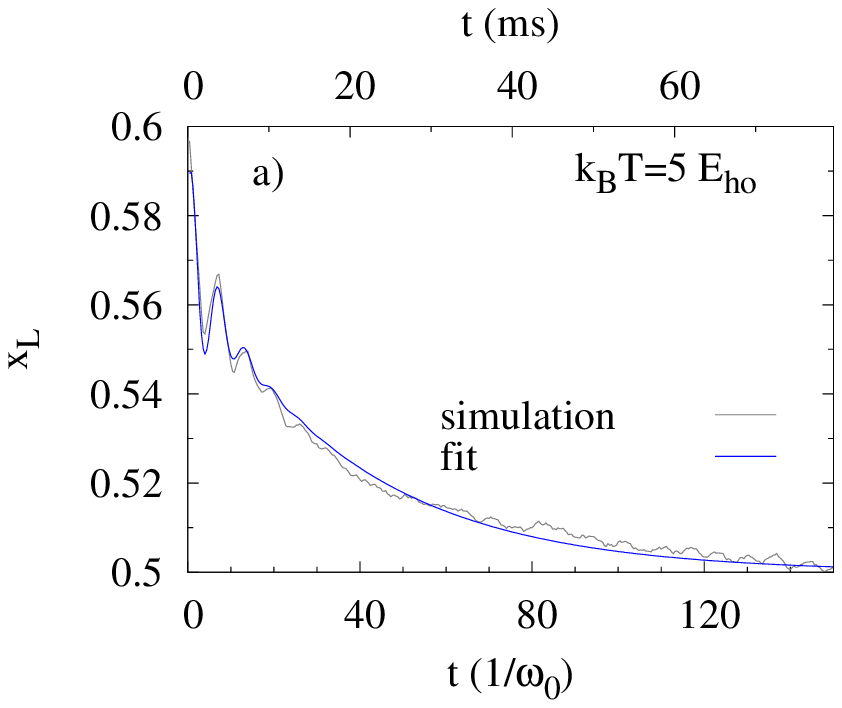}
\includegraphics[width=5.9cm,angle=0]{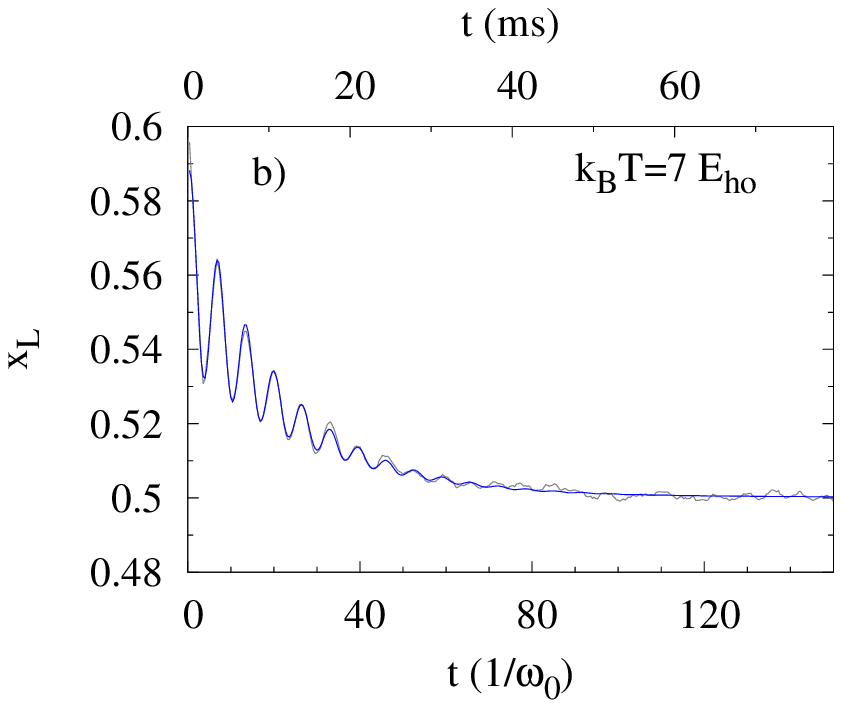}
\includegraphics[width=5.9cm,angle=0]{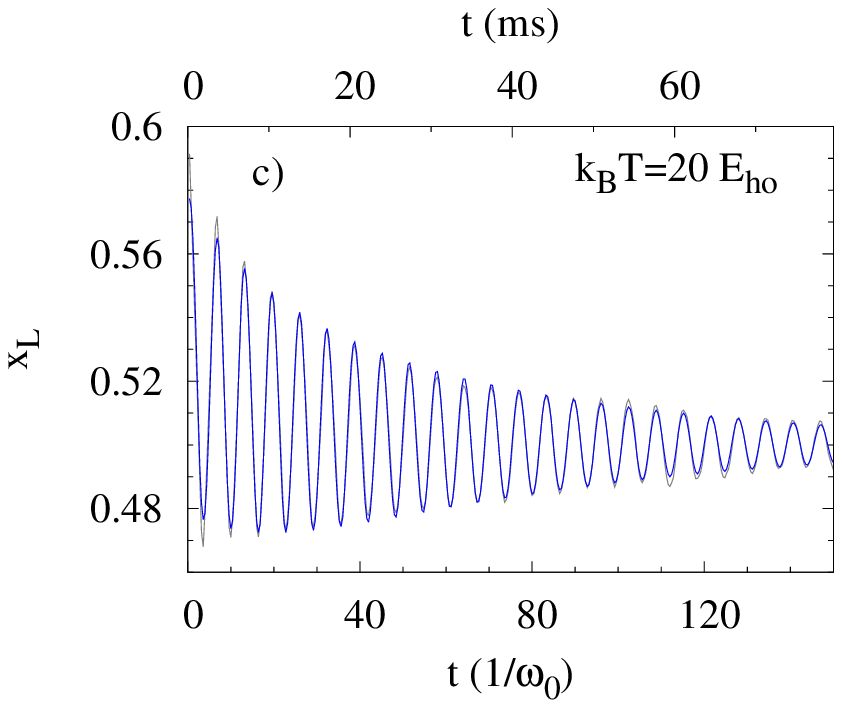}
\end{center}
\caption{Dynamics in the HGDW. Evolution of $x_L$ with time, for three 
values of the temperature. The gas contains $N=5000$ particles, with an 
initial imbalance of $x_L(t=0)=0.6$. 
The well parameters, in trap units, are 
$\tilde{V}=10$, $d_{int}=0.06$, $w=0.8$ (therefore $V_0\simeq 7.6$), 
and the temperatures $T=5$ (panel {\it a}), $7$ (panel {\it b}) 
and $20$ (panel {\it c}). 
To express them in dimensional units, reported 
in the top $x$-axis, we need to specify the trap frequency 
$\omega_0$ and the particle mass: 
for $^{6}$Li and $\omega_0=2\pi\times 300$ Hz they correspond to $\tilde{V}/k_B=144nK$, 
$d_{int}=142 nm=2684 a_0$, $w= 1.89 \mu m$, $T=72, 101, 288 nK$. 
The values of $\tau$ and $\omega$ fitted from the data 
for figures ({\it a,b,c}) in trap units are respectively  
$\tau=(37.2\pm0.2, 22.8\pm0.2, 18.6\pm0.4)$ and 
$\omega=(0.99\pm0.01, 0.962\pm0.002, 0.9859\pm0.0001)$.
}
\label{fig:xLvstReal}
\end{figure*}

In \Fig{fig:xLvstReal} 
we show three typical behaviours of $x_L(t)$: they correspond to a given 
cross section and different temperatures, increasing
from left to right.
The frequency $\omega$ of the c.o.m. oscillation is very close to 
$\omega_0$ at high temperatures, 
where the presence of the barrier does almost not 
affect the cloud oscillation; 
at low temperatures, instead, it is reduced to smaller values. 

\begin{figure}[t]
\begin{center}
\includegraphics[height=7.5cm,angle=0]{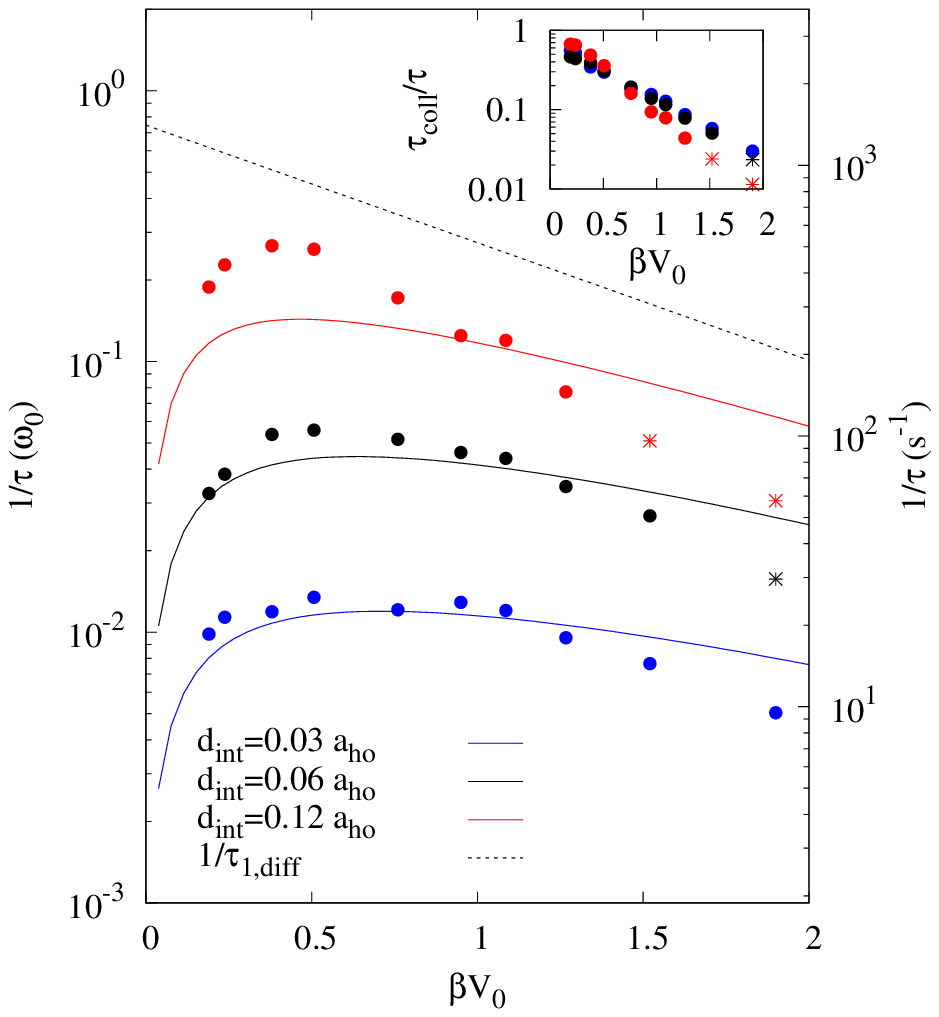}
\end{center}
\caption{HGDW relaxation time. The inverse relaxation time $1/\tau$ as a function of
 $\beta V_0$, obtained varying the temperature 
(the trap is fixed as in \Fig{fig:xLvstReal} with a 
barrier height $V_0\simeq 7.6$).
The numerical results are denoted by points. 
Different colors correspond to different values of the cross section:
$\sigma=\pi \times(0.03, 0.06, 0.12)^2=\pi\times(71,142,284)^2\text{nm}^2$.
Each point is obtained from an average over $40$ runs. 
The fits are done in the whole time interval, using either
$f_2(t)$ (full circles), or
$f_3(t)$ (stars).
The dimensional 
quantities reported in the figure are obtained as in \Fig{fig:xLvstReal}. 
For comparison, we show as solid lines the results obtained in 
an approximated analytical solution presented in \Sec{sec:modelLowBarr}, 
namely $\tau_1$ obtained 
using Eqs. (\ref{eq:tau1}), (\ref{eq:xaboeq}), (\ref{eq:ksreal1}) and (\ref{eq:ksreal2}).
Inset: $\tau$ rescaled in units of collisional time $\tau_\coll$.  
Colors correspond to the same cross section as the main figure.
}
\label{fig:tauVSbetaV0real}
\end{figure}

Repeating similar calculations for a set of temperatures and interaction 
strengths, we obtain the results shown in \Fig{fig:tauVSbetaV0real}.
In some cases, in which the c.o.m. oscillation is not visible anymore, 
we use the fitting function $f_3$ to extract the relaxation time. 
As before, the symbol shape indicates the fitting function used, with the 
same notation of \Fig{fig:tauVSbetaV0}. 
Comparing with the analogous plot for the SDW, \Fig{fig:tauVSbetaV0},
we see an analogous accumulation of curves as the cross section increases.
The dependence upon $\beta V_0$ seems qualitatively different: 
the point is that here, along a curve for a given cross section, the global collision rate changes, whereas 
this was not the case in \Fig{fig:tauVSbetaV0}. If we rescale all the 
curves by the corresponding equilibrium collision rate (inset), 
we see that they all fall in the same region and the trend so obtained 
is therefore similar in SDW and the HGDW.

In panel {\it a)} of \Fig{fig:tauVSbetaV0real}, 
together with the numerical results, we show 
also (full lines) the analytical predictions obtained extending our model 
to the realistic well case (see \Sec{sec:adapt}).

\begin{figure}[t]
\begin{center}
\includegraphics[width=8cm,angle=0]{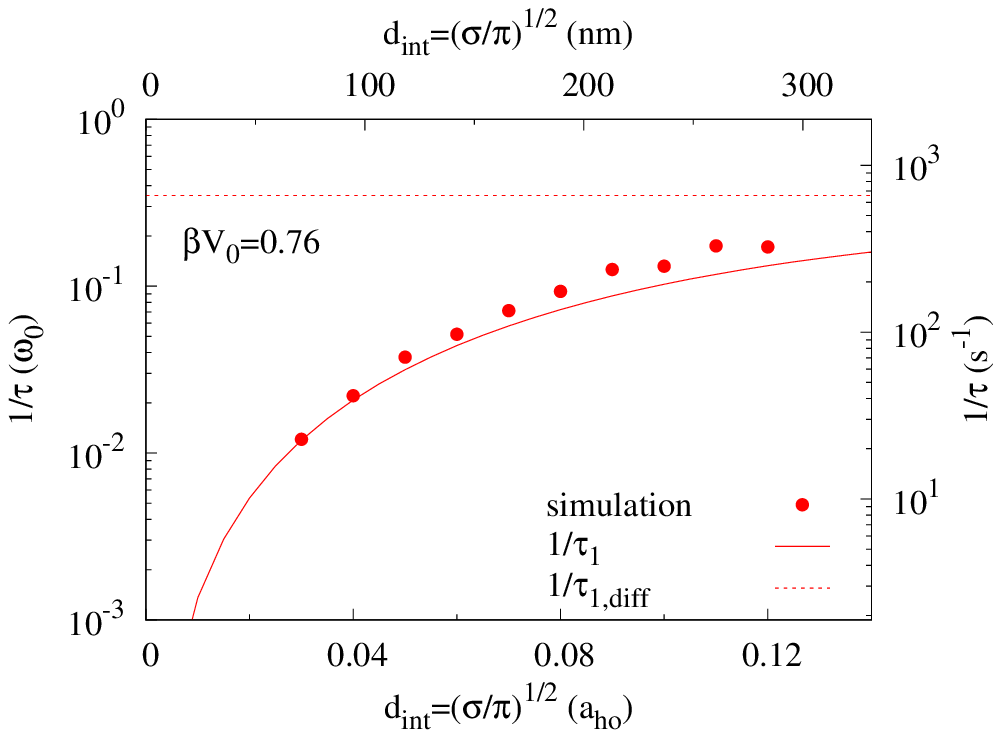}
\end{center}
\caption{HGDW potential. Inverse relaxation time $1/\tau$
 as a function of 
$d_\inter$ for the same trap of \Fig{fig:xLvstReal} and a given temperature
 ($T=10$, then $\beta V_0=0.76$). 
The fitting function and its parameters are the same as in 
\Fig{fig:xLvstReal}.
}
\label{fig:tauVSdreal}
\end{figure}

Next, in \Fig{fig:tauVSdreal},
we show the analogous of \Fig{fig:tauVSd} for the realistic HGDW well,
 for $\beta V_0\simeq 0.76$: the qualitative behaviour is the same and 
it is nicely reproduced by the analytical 
model.

In the figures of this Section 
(\Fig{fig:xLvstReal}, \Fig{fig:tauVSbetaV0real}, \Fig{fig:tauVSdreal}), 
together 
with the harmonic oscillator units, we show also axes with physical units: they 
are obtained assuming a reference mass (that of $^6Li$) and a 
trap frequency $\omega_0=2\pi\times 300$ Hz.

\subsubsection{Estimates for a two-component Fermi gas}

\begin{figure}[t]
\begin{center}
\includegraphics[width=8cm,angle=0]{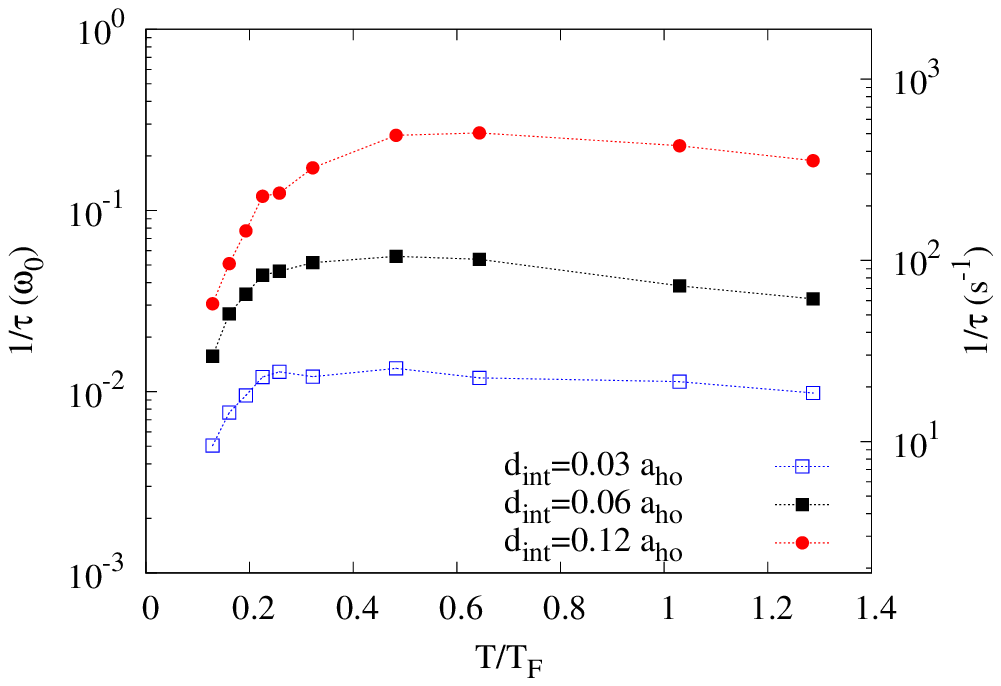}
\end{center}
\caption{HGDW potential. Inverse relaxation time $1/\tau$
 as a function of $T/T_F$, for a fixed barrier $V_0\simeq 7.6$. 
The results for $\tau$ are the same of \Fig{fig:tauVSbetaV0real}, here shown in 
a different representation. 
The Fermi temperature has been computed as 
$T_F=(6 N_\uparrow)^{1/3}$: that is, the Fermi temperature for a balanced mixture 
having $N_\uparrow=N_\downarrow=N$ in a harmonic well. 
For details on the parameters and the dimensional units, see the caption of 
\Fig{fig:xLvstReal}.
}
\label{fig:tauVST2TF}
\end{figure}

Finally, we can use our results to make an approximate prediction for a 
balanced two-component mixture of $^6Li$. In fact, in this case we would have 
two species, equally populated ($N_\uparrow=N_\downarrow$) with only 
inter-species interactions. In the Boltzmann framework, we would have 
two distribution functions, but they coincide if the mode and the 
potential do not depend upon the ``spin'': just one 
distribution normalized to $N_\uparrow$ is needed. Our calculations 
for $5000$ classical particles 
correspond therefore to a balanced mixture of $10^4$ fermions.
Anyway, notice that we are not including Pauli blocking in the 
collision term and we are approximating the cross section with a 
constant. With this in mind, in \Fig{fig:tauVST2TF}
we plot the same results of \Fig{fig:tauVSbetaV0real}, as a function 
of $T/T_F$, with $T_F=(6 N_\uparrow)^{1/3}$
the Fermi temperature in a harmonic trap, and $N_\uparrow=5000$. 
Notice that even we are not considering 
the effect of the quantum statistics on the collision term, 
both numerical computations and the 
analytical model can be straightforwardly extended 
to include it: we expect that, at temperatures $T/T_F \gtrsim 0.5$ (and in any 
case larger than $T_c$), the 
relaxation time has a similar qualitative dependence on 
barrier heights and interaction strengths.

\section{Conclusions}\label{sec:concl}

In this work we have presented a study of the double-well 
dynamics of a classical gas that obeys the Boltzmann equation, 
with the purpose of assessing the role played by collisions in
the relaxation of population imbalance. We think that 
a detailed study of the Boltzmann equation in a variety of double-well 
potentials is a rewarding subject of interest, not only 
for its paradigmatic and pedagogical importance, but also 
to concretely model currently ongoing experiments in a range of temperature 
$T \gtrsim T_c$ and to set a basis for further 
quantitative theoretical studies  
of tunneling of ultracold atoms at finite temperature below $T_c$, 
in particular near $T_c$. 

Two model potentials have been considered: a toy square double-well 
(SDW) potential with a filtering wall and a more realistic double-well 
obtained by combining a gaussian to a harmonic potential (HGDW). 
In both cases, we have performed numerical simulations (test-particle method) 
of the relaxation dynamics from an initial imbalanced population 
of the symmetric wells to the balanced equilibrium, 
in a range of cross sections $\sigma$ values
and barrier heights $V_0/T$.
For convenience, in the toy SDW 
potential we have fixed the temperature and varied the 
barrier height, whereas in the realistic HGDW 
one we have fixed the trap shape, therefore 
the barrier height, and varied the temperature.

Beside the numerical results, we have also presented a simple analytical model 
for the dynamics, that allows to compute the relaxation time analytically from 
the system parameters. The agreement of the model predictions 
with the results of the simulations is qualitative, and it is 
quantitatively better for the realistic HGDW potential. In general 
the analytical findings are in reasonable agreement with the 
numerical results for low up to intermediate barriers ($\beta V_0 \lesssim 1$).

Finally, we have used our results to estimate the relaxation times 
for realistic values of on-going experiments 
with a mixture of fermionic cold atoms ($^6Li$).

As a possible continuation of this study, it would be interesting to compare 
our results with those obtained in the framework of Klein-Kramers equation 
in presence of a barrier \cite{Hanggi90}: to this end, the friction parameter 
appearing in the Klein-Kramers equation should be appropriately (we mean, 
quantitatively) computed.

Another extension of this work, relevant for ultracold atom experiments 
at low temperature, would be to include, beyond 
the classical crossing mechanism studied here, the hopping 
via quantum tunneling: a possible way could be to include 
an effective {\it ad hoc} term in the 
Boltzmann equation or to couple the Boltzmann equation to the equations 
for the superfluid dynamics.

\section*{Acknowledgments}

S.C. is supported by the 
``Funda\c{c}\~ao para a Ci\^encia e a Tecnologia'' (FCT, Portugal) and
the ``European Social Fund'' (ESF) via the post-doctoral grant
SFRH/BPD/64405/2009. A.T. acknowledges support from the
Italian PRIN ``Fenomeni quantistici collettivi: dai sistemi fortemente 
correlati ai simulatori quantistici'' (PRIN 2010\_2010LLKJBX).

The authors are grateful to D. Davesne, A. Gambassi, A. Laio 
and I. Vida\~{n}a for valuable 
discussions. We also gratefully thank for many useful discussions 
people of LENS and QSTAR groups in Florence, in particular 
A. Burchianti, K. Xhani, 
G. Roati, A. Smerzi and M. Zaccanti. 
A.T. acknowledges the University of Coimbra 
for kind hospitality and the Isaac Newton Institute for Mathematical Sciences, 
Programme ``Mathematical Aspects of Quantum Integrable Models in and out of Equilibrium'', 
where the final part of this work was completed. 
S.C. acknowledges CNR-IOM for kind hospitality in Trieste. 
The authors also thank
the Laboratory for Advanced Computing at the University of Coimbra 
for providing CPU time in the Milipeia and Navigator clusters. 

\appendix

\section{Collision rates at equilibrium}\label{app:rates}

In a generic potential and for constant cross-section, the collision rate 
at equilibrium reads
\begin{equation}
\Gamma_{eq}=\frac{2\sigma}{\sqrt{\pi\beta m}}\int d^3 r \rho_\eq^2({\bf r})~,
\end{equation}
where $\rho_\eq({\bf r})=\int d^3p/(2\pi\hbar)^3 f_\eq({\bf r},{\bf p})$ 
is the local equilibrium density: in some cases, this integral can 
be performed analytically.

For a box with hard walls, the distribution
function reads
\begin{equation}
f_{\eq}(\vek{p})=e^{-\beta\left(\frac{p^2}{2m}-\mu\right)}~,
\end{equation}
where $\mu$ the chemical potential. 
The normalization condition $\int d\Gamma f = N$ 
sets the value of the chemical potential $e^{\beta\mu}=\hbar^3(2\pi\beta/m)^{3/2} 
\rho$ where $\rho=N/\Omega$. One then finds
\begin{equation}\label{eq:Gammaeqbox}
\Gamma_\eq^{\mathit box}=\frac{2 N^2 \sigma}{\Omega\sqrt{\pi\beta m}}= 
\frac{2 N {\rho}\sigma}{\sqrt{\pi\beta m}}~,
\end{equation}
where we have used in the last equality $\rho=N/\Omega$. 

The average time in a square single-well between 
two consecutive collisions of the same particle 
is 
\begin{equation}
\tau_{\mathit coll}^{\mathit box}= (2\Gamma_\eq^{\mathit box}/N)^{-1}=\frac{1}{4}\frac{\sqrt{\pi\beta m}}
{\rho\sigma}~.
\end{equation}
On the other hand, the average time a particle needs to travel across the whole 
box is
\begin{equation}
\tau_{\mathit trap}^{\mathit box}= \frac{\Omega^{1/3}}{v_{\rms}} = \Omega^{1/3}\sqrt{\frac{\beta m}{3}}~,
\end{equation}
where the root-mean-square velocity at equilibrium is 
$v_\rms=\sqrt{3 /(\beta m)}$.
Defining the adimensional quantity 
$\alpha \equiv \tau_{\mathit coll}^{\mathit box}/\tau_{\mathit trap}^{\mathit box}$,
if $\alpha\rightarrow 0$ the gas is hydrodynamical, whereas it is 
collision-less if $\alpha \rightarrow \infty$. The condition on $\alpha$ 
turns out to be a purely geometrical one since 
\begin{equation}
\alpha=\frac{\sqrt{3\pi}}{4} \frac{1}{\Omega^{1/3}\rho \sigma}:
\end{equation}
if the volume and number of particles are fixed, it is a condition on the 
cross section $\sigma$.

Finally, we use in the text the the 
equilibrium collision rate for a harmonic anisotropic trap 
$V({\bf r})=m(\omega_x^2x^2+\omega_y^2y^2+\omega_z^2z^2)/2$, 
which is given by
\begin{equation}
\Gamma_\eq^{HO}=\frac{\sigma N^2\beta m \omega_x\omega_y\omega_z}{4 \pi^2}~.
\end{equation}
%

\section{Chemical potentials for the square single-well}\label{app:chem}

Regarding the chemical potential, in a square single-well $\mu$ is such that
\begin{equation}
e^{-\beta\mu}=\frac{1}{N}\frac{\Omega}{(2\pi\beta/m)^{3/2}\hbar^3}~.
\end{equation} 
Therefore the chemical potentials $\mu^\gtrless$ entering 
the analytical model presented in \Sec{sec:modelLowBarr} are given by
\begin{equation}
e^{-\beta\mu^<}=\frac{\Omega}{(2\pi\beta/m)^{3/2}\hbar^3} ~\erfB
\end{equation}
and 
\begin{equation}
e^{-\beta\mu^>}=\frac{\Omega}{(2\pi\beta/m)^{3/2}\hbar^3} ~\erfcB~.
\end{equation}
The equilibrium values of $N^\gtrless$ are
\begin{equation}
N^>_\eq=N \erfcB~,\quad  N^<_\eq=N \erfB~.
\end{equation}
%

\section{Rate coefficients $\gamma_i$ and $h_i$}\label{sec:h1h2h3}

For a gas of $N$ particles in a volume $\Omega$, the rates introduced in \Sec{sec:singletoy}
(see relations \ref{ratedef}, \ref{eq:Nupdot}, \ref{eq:Gammas}) are
\begin{widetext}
\begin{equation} \label{gamma123_px}
\left\{
\begin{array}{ccl}
\gamma_1 &=& \Gamma_A^\eq = e^{2\beta\mu} ~\Omega \int \frac{d^3p_1 d^3p_2}{(2\pi\hbar)^6}\ \frac{d\sigma}{d\Omega'}d\Omega' \
\frac{|p_1-p_2|}{m} e^{-\beta \frac{p_1^2+p_2^2}{2m}} \Theta(p_{1x}^2-p_0^2)  \Theta(p_0^2-p_{2x}^2)
 \Theta(p_0^2-{{p'}}_{1x}^2) \Theta(p_0^2-{p'}_{2x}^2)\\
\gamma_2 &=& \Gamma_C^\eq =  e^{2\beta\mu}~ \Omega \int \frac{d^3p_1 d^3p_2}{(2\pi\hbar)^6}\ \frac{d\sigma}{d\Omega'}d\Omega' \
\frac{|p_1-p_2|}{m} e^{-\beta \frac{p_1^2+p_2^2}{2m}} \Theta(p_0^2-p_{1x}^2)  \Theta(p_{2x}^2-p_0^2)
 \Theta({p'}_{1x}^2-p_0^2) \Theta({p'}_{2x}^2-p_0^2)\\
\gamma_3 &=& 2\Gamma_E^\eq = e^{2\beta\mu} ~ \Omega \int \frac{d^3p_1 d^3p_2}{(2\pi\hbar)^6}\ \frac{d\sigma}{d\Omega'}d\Omega' \
\frac{|p_1-p_2|}{m} e^{-\beta \frac{p_1^2+p_2^2}{2m}} \Theta(p_0^2-p_{1x}^2)  \Theta(p_0^2-p_{2x}^2)
 \Theta({p'}_{1x}^2-p_0^2) \Theta({p'}_{2x}^2-p_0^2)~,
\end{array} \right.
\end{equation}
\end{widetext}
where the prefactors come from the relation
$e^{\beta\mu_>}N^>_\eq= e^{\beta\mu_<}N^<_\eq =e^{\beta\mu}$ that can 
be easily checked.
The equilibrium total collision rate is
\begin{equation}
\Gamma_\eq = \frac{1}{2}e^{2\beta\mu} ~\Omega \int \frac{d^3p_1 d^3p_2}{(2\pi\hbar)^6}\ \frac{d\sigma}{d\Omega'}d\Omega' \
\frac{|p_1-p_2|}{m} e^{-\beta \frac{p_1^2+p_2^2}{2m}}~.
\end{equation}
By passing to adimensional variables, it's easy to see that 
$h_i\equiv \gamma_i/\Gamma_\eq$ are functions of 
$\beta p_0^2/2m$ (i. e., of $\beta V_0$) only.

Notice that these ratios are global equilibrium quantities. We computed them 
in the uniform case, however they are 
{\it the same} in any potential $V(\vek{r})$: in fact, the spatial 
dependence would factorize out and cancel between numerator and 
denominator.


\end{document}